\documentclass[aps,pre,superscriptaddress,showpacs,twocolumns]{revtex4}
\usepackage{graphicx,color,amsmath,amssymb}
\usepackage{dcolumn}
\usepackage{bm}				
\usepackage{subfigure}

\begin{document}

\title{Laterally driven interfaces in the three-dimensional Ising lattice gas}

\author{Thomas H. R. Smith}
\affiliation{H.H. Wills Physics Laboratory, University of Bristol,
  					 Tyndall Avenue, Bristol BS8 1TL, United Kingdom}
\author{Oleg Vasilyev}
\affiliation{Max-Planck-Institut f{\"u}r Metallforschung,
  					 Heisenbergstra\ss e~3, D-70569 Stuttgart, Germany}
\affiliation{Institut f{\"u}r Theoretische und Angewandte Physik,
             Universit{\"a}t Stuttgart, Pfaffenwaldring 57,
             D-70569 Stuttgart, Germany}
\author{Anna Macio\l ek}
\affiliation{Max-Planck-Institut f{\"u}r Metallforschung,
  					 Heisenbergstra\ss e~3, D-70569 Stuttgart, Germany}
\affiliation{Institut f{\"u}r Theoretische und Angewandte Physik,
             Universit{\"a}t Stuttgart, Pfaffenwaldring 57,
             D-70569 Stuttgart, Germany}
\affiliation{Institute of Physical Chemistry, Polish Academy of Sciences,
						 Department III, Kasprzaka 44/52, PL-01-224 Warsaw, Poland}
\author{Matthias Schmidt}
\affiliation{H.H. Wills Physics Laboratory, University of Bristol,
  					 Tyndall Avenue, Bristol BS8 1TL, United Kingdom}
\affiliation{Theoretische Physik II, Universit\"at Bayreuth, Universit\"atsstra\ss e
  					 30,  D-95440 Bayreuth, Germany}
\date{\today}

\begin{abstract}
We study the steady state of a phase-separated driven Ising lattice gas in three dimensions using
computer simulations with Kawasaki dynamics. An external force field ${\bf F}(z)$ acts in the $x$
direction parallel to the interface, creating a lateral order parameter current $j^x(z)$ which
varies with distance $z$ from the interface. Above the roughening temperature, our data for
`shear-like' linear variation of ${\bf F}(z)$ are in agreement with the picture wherein  shear acts
as effective confinement in this system, thus supressing the interfacial capillary-wave
fluctuations. We find sharper magnetisation profiles and reduced interfacial width as compared to
equilibrium. Pair correlations are  more suppressed in the vorticity direction $y$ than in the
driving direction; the opposite holds for the structure factor. Lateral transport of capillary waves
occurs for those forms of ${\bf F}(z)$ for which the current $j^x(z)$ is an odd function of $z$, for
example the shear-like drive, and a `step-like' driving field.  For a V-shaped driving force no
such motion occurs, but capillary waves are suppressed more strongly than for the shear-like drive.
These findings are in agreement with our previous simulation studies in two dimensions.
Near and below the (equilibrium) roughening temperature the effective-confinement picture ceases to
work, but the lateral motion of the interface persists.
\end{abstract}

\pacs{05.40.-a, 05.50.+q, 68.05.Cf, 68.35.Rh}

\maketitle

\section{Introduction}

Dimensionality $d$ of space is a relevant parameter in condensed matter systems that exhibit large
spatial fluctuations, e.g., thermal fluctuations of the order parameter near a second-order phase
transition. Such critical fluctuations are correlated over large distances, which gives rise to
singular behaviour characterized by a set of critical exponents whose values depend on $d$.
Moreover, the very existence of the phase transition depends on $d$. Below the lower critical
dimension fluctuations are strong enough to destroy the ordered phase, and hence there is no phase
transition. In contrast, above the upper critical dimension fluctuations are no longer important,
and the critical exponents become the same as in mean field theory.

An important dependence on dimensionality also occurs in systems where two distinct phases coexist.
In such systems thermal fluctuations can be correlated over a long distance {\it within} the
interfacial region. Then the interfacial correlation length $\xi_{\parallel}$ (parallel to the
interface) increases with increasing thickness $w$ of the interface, $\xi_{\parallel}\propto
w^{\zeta}$, where $0\le \zeta < 1$ \cite{Fisher}, and the interface is termed {\it rough}. The value
of the roughness exponent $\zeta$ again depends on dimensionality $d$. A statistical-mechanical
description of interfacial degrees of freedom, capillary wave theory, was proposed by Buff, Lovett
and Stillinger \cite{BLS,BedeauxWeeks}, who suggested to model the interface as a sharp divide
between the two phases, but one that would freely fluctuate. Capillary wave theory (CWT) predicts
that $\zeta=\frac{1}{2}(3-d)$ for $d\le 3$, i.e., fluctuations can destabilise the interface in $d
\le 3$. Then the thickness $w$ of the free interface is infinite in the thermodynamic limit. At the
marginal dimension $d=3$, the value $\zeta=0$ corresponds to a logarithmic divergence. For $d>3$,
one has $\zeta=0$ but $w$ is finite for all $\xi_{\parallel}$; the interface is then said to be
smooth.

These results originate from the choice of the simplest, Gaussian form of fluctuations, i.e., the
probability for a local departure (height) $h(\mathbf{r})$ of the interface from the reference plane
$h=0$ is given by the Boltzmann weight $\propto e^{-\Delta {\cal H} / k_B T}$, where $k_BT$ is the
thermal energy, and

\begin{equation}
\label{eq:0}
\Delta {\cal H} = \int d^{d-1}{\bf r} \left( \frac{\Gamma}{2} \left[\nabla h({\bf r)}\right]^2 + 
									V_{\rm ext}[h({\bf r})] \right),
\end{equation}
where $\mathbf{r}$ are coordinates parallel to the interface. $\Gamma$ is the interfacial stiffness,
which for a continuum fluid is simply the tension of a free interface
\cite{Fisher,LipowskyFisher,Jasnow,Tarazona}. If the external potential $V_{\rm ext}$ is quadratic
in $h(\mathbf{r})$, for example due to gravity, the equipartition theorem for quadratic degrees of
freedom may be applied. The equilibrium interface pair correlation function for a
translationally-invariant system is then found to be
\begin{equation}
\label{eq:hcorr}
C({\mathbf r}) \equiv \left< h(0)h({\mathbf r}) \right> = 
		 \frac{k_BT}{\Gamma}
		 \int \frac{d^{d-1}q}{(2\pi)^{d-1}}
		 \frac{e^{i \mathbf{q} \cdot \mathbf{r}}}{\xi_{\parallel}^{-2}+q^2},
\end{equation}
where $\xi_{\parallel}^{-2}=\Gamma^{-1}(\partial^2 V_{\rm ext}/\partial h^2)$, and the limit of infinite
interface dimension $L^{d-1}$, $L\to \infty$ has been taken. An upper cutoff on wave numbers
$\left|{\mathbf q}\right| \le \pi / a$ is always assumed. $a$ is usually identified with some
appropriate microscopic length in the interface region, e.g., the lattice spacing or the bulk
correlation length $\xi_b$ \cite{Weeks77}. The behaviour of $C({\mathbf r})$ is obviously
$d$-dependent -- as is the interface width $w$, defined by $w^2 = C(0)$.

For certain microscopic models, such as the Ising model, which is equivalent to the lattice gas
model of a fluid, exact results are available \cite{DBA_PTCP,Jasnow}. A description starting from a
microscopic Hamiltonian is particularly useful because it accounts for both interfacial and bulk
fluctuations. In contrast bulk degrees of freedom of coexisting phases are absent in CWT, in which
one considers only interfacial configurations. On the other hand, in the ``classical'' theories for
the interface based on order parameter (or density) profiles, for example the van der Waals theory
\cite{vdW,RowlinsonandWidom}, interfacial fluctuations are absent and the structure of the interface
is reduced to an inhomogeneity of the order parameter, i.e., it is of order of the bulk correlation
length $\xi_b$.

In $d=2$, Ising interfaces are rough for all temperatures $T$ below the critical temperature $T_c$,
and $w^2\propto L$, as revealed by exact results for the local magnetisation profile on a scale
\emph{large} compared to $\xi_b$ \cite{Douglas}, in agreement with CWT. However, in $d=3$ there is
evidence \cite{Jasnow,WGL,Weeks80} for the existence of a finite roughening temperature $0<T_R<T_c$,
below which the interface is smooth. This evidence is supported by Monte Carlo (MC) simulations
\cite{HP} and by rigorous analysis of discrete random surface models, such as the solid-on-solid
(SOS) and discrete Gaussian (DG) models \cite{Weeks80,Whiskers}. These models approximate the
interface of the Ising model at low temperatures, for example the SOS model can be regarded as a
certain anisotropic-coupling limit of the Ising model where the SOS interface configurations are 
selected from the Ising configurations by the requirement that there are no overhangs or bubbles
\cite{DBA_PTCP}.

In the situations that we shall address, the interface is stabilised by the presence of two walls at
spacing $L_{\perp}$. This problem is interesting because the fluctuating interface will experience
collisions with the constraining walls. The standard capillary wave model does not apply for this
case; it has to be extended to take into account entropic repulsion from the walls. Confined Ising
interfaces can be treated rigorously in $d=2$. Results for the magnetisation profile \cite{Maciolek}
indicate that in two dimensions, in spite of the entropic repulsion at its extremities, the
interface meanders back and forth between the walls so that $w~\propto~L_{\perp}$. This is markedly
different from what is expected in three dimensions on the basis of the analysis of low-energy
excitations in discrete random surface models \cite{Whiskers}. An energy-versus-entropy argument
leads to the conclusion that interface configurations which result in interference with the boundary
are needle-like. A rigorous analysis gives $w^2 \propto L_{\perp}$ and $L_{\perp} \propto \ln
\xi_{\parallel}$. These results support conjectures from the phenomenological effective interface
Hamiltonian \cite{Fisher,FandF,LipowskyFisher}, and for $d=3$ Ising interfaces they agree with MC
simulation studies \cite{KKB}.

Dependence of the structure of equilibrium interfaces on the spatial dimensionality $d$ has
repercussions also on relaxation dynamics of the fluctuating interface. Relaxation dynamics of
capillary waves have been recently studied using molecular dynamics simulations of simple liquids in
$d=3$ \cite{Tarazona,Thakre}. For the liquid interface constrained between two walls at separation
$L_{\perp}$, a pronounced enhancement of the relaxation time of capillary waves was found; the most
affected are relaxation times of waves with the wave number $q\simeq L_{\perp}^{-1}$ \cite{Thakre}.

These results for equilibrium suggest that a non-trivial dependence of the interfacial structure and
dynamics on dimensionality may persist to non-equilibrium situations. In this paper, we study this
problem for fluctuating interfaces that are driven into a steady state by the action of an external
field parallel to the plane of the interface. The problems of roughness, spatial and temporal
correlations of laterally driven interfaces were first addressed in the lattice gas driven by a
spatially uniform force field via MC simulations \cite{LeungMon89,LeungZia93}. These studies, which
were of two-dimensional systems, found that the interface becomes less rough when drive is applied.
The order-parameter (magnetisation) profiles become much sharper upon increasing the drive and the
interfacial width is reduced as compared to the equilibrium value. By a suitable coarse graining of
microscopic particle configurations, the local position (height) of the interface was defined, and
the behaviour of the spatial interface height correlation function was studied. The results are
consistent with a reduction of the interfacial correlation length $\xi_{\parallel}$. Moreover, the
structure factor $S(q)$, defined as the the Fourier transform of the height-height correlation
function displays deviations from the equilibrium capillary wave dependence $1/q^2$ as $q\to 0$; the
data suggest a weaker singularity $S(q)\propto 1/q^{0.67}$, which implies a reduction of the
roughness exponent $\zeta$. In theoretical  attempts to treat driven interfaces, one derives a
dynamic equation for the interfacial degrees of freedom, starting from the time-dependent
Landau-Ginzburg-type equation for the order parameter \cite{Leung88,ZiaLeung91}. This approach leads
to a non-local and non-linear equation for the interface height. A linear stability analysis of this
equation for a spatially uniform drive parallel to the interface shows that temporal decay of
fluctuations along the driving field is faster than that orthogonal to the driving field. However,
predictions for the roughness of the interface do not agree with the simulation results in $d=2$.

More recent interest in the theoretical challenges of laterally driven interfaces originates from an
experiment on colloidal gas-liquid interfaces subjected to shear flow. In the experiment of Derks
\emph{et al.\ }\cite{DerksShear} a real-space visualisation of interfacial fluctuations revealed a
reduced interfacial roughness when shear was applied. The width $w$  and correlation function of the
height of the interface were analysed by fitting to the equilibrium CWT results. The fitting
parameters in the analytic CWT expressions were the correlation length $\xi_{\parallel}$ along the
interface measured in the flow direction, and the surface tension. The authors concluded that $w$
was decreased but $\xi_{\parallel}$ was increased. Recently, the problem of non-equilibrium
fluctuations of a liquid-liquid interface under shear has been addressed theoretically
\cite{ThieBick} within the framework of fluctuating hydrodynamics. This approach leads to a
mode-coupling equation for the interface height which was solved using a perturbation theory.
Results for the interfacial width are in agreement with the experiment of Ref.~\cite{DerksShear},
but the results for the interfacial correlation length in the flow direction are not. The
theoretical height-height correlation function and the structure factor imply a decrease of the
correlation length in the direction of flow. Interestingly, in the direction perpendicular to the
flow (vorticity direction), the correlation length seems to \emph{increase}. Similar conclusions
have been obtained from molecular dynamics simulations \cite{Thakre}.

Previously we have studied  interfaces in the two-dimensional Ising strip driven by an external
field that is applied parallel to the walls (and to the interface), and may vary in the direction
perpendicular to the mean position of the interface, by using MC simulations with spin-exchange
Kawasaki dynamics \cite{Shear2dPaper,LM7Proc,MovPaper}. These studies were partially motivated by
the need to understand sheared fluid interfaces. Because our results were obtained in $d=2$, and
because of the simplified character of our model and its dynamics, we were not in the position to
attempt a direct comparison with experimental data. However, our results were in \emph{partial}
qualitative agreement with Ref.~\cite{DerksShear}. We found that the shear-like drive acts as an
effective confinement on the system; a steady state is reached in which the magnetisation profile is
the same as that in equilibrium, but with a rescaled length implying a reduction of the interfacial
width. Pair correlation functions along the interface decay more rapidly with distance under drive
than in equilibrium, and for cases of weak drive can be rescaled to the equilibrium result.
Moreover, we find that interfacial transport can occur in an unexpected way parallel to the
interface. The lateral flux of the order parameter at a planar interface induces lateral motion of
the thermal capillary waves, provided that the flux is an odd function of distance from the
interface. 

In the present paper we study the same model system using the same approach, but in three spatial
dimensions. We wish to investigate to what extent our findings from $2d$ persist to higher
dimensions. Moreover, dependence on dimensionality of various quantities characterising the
structure and dynamics of the interface, as well as of transport properties in a driven state is
interesting. One would like to know whether the different character of interfacial fluctuations,
i.e., ``wandering'' in $d=2$ and ``spikes'' \cite{Whiskers} in $d=3$ has any repercussions. There
are also new aspects that deserve to be studied: the behaviour of the system near the roughening
transition, and the anisotropy effects introduced by the introduction of the vorticity direction.
Last but not least, results in $d=3$ permit a more direct comparison to experiment.

The rest of this paper is organised as follows. In Section \ref{sec:model} we introduce the model
and give details of the simulations. In Section \ref{sec:static_results}, we present and discuss
results for the interfacial structure of the driven $3d$ Ising system, via the magnetisation
profile, interface width, and correlation functions in real and Fourier space. In Section
\ref{sec:dynamic_results}, we investigate the dynamics of the driven interface, showing results for
the current and evidence for capillary wave transport. Finally we draw conclusions in Section
\ref{sec:concout}.

%%%%%%%%%%%%%%%%%%%%%%%%%%%%					MODEL/SIMULATION  			%%%%%%%%%%%%%%%%%%%%%%%%%%%%%%%%%%%%%%
%%%%%%%%%%%%%%%%%%%%%%%%%%%%%%%%%%%%%%%%%%%%%%%%%%%%%%%%%%%%%%%%%%%%%%%%%%%%%%%%%%%%%%%%%%%%%%%%%%%%

\section{Model and simulation details} \label{sec:model}

We consider a three-dimensional (3\emph{d}) Ising model on a simple cubic lattice. On lattice sites
$i$ sit spins $\sigma_i$, which may take values $\pm 1$. In lattice gas language, the spin variables
become particle occupation numbers $\tau_i = (\sigma + 1)/2 = 0,1$, corresponding to the absence and
presence of a particle at a site, respectively. The Hamiltonian for the system is
\begin{equation}
\label{eq:hamiltonian}
H = -J \sum_{\left< i,j \right>}\sigma_i \sigma_j,
\end{equation}
where $\left< i,j \right>$ indicates that the sum is over nearest neighbour sites $i$ and $j$. The
coupling constant $J > 0$, so that the interactions are ferromagnetic (attractive in lattice gas
language).

The lattice has dimensions $L_x \times L_y \times L_z$, with periodic boundary conditions applied in
the $x$ and $y$ directions. (All lengths are expressed in units of the lattice constant $b=1$). Spin
layers are located at heights $z=-\frac{L_z - 1}{2},-\frac{L_z - 3}{2} \dots,-\frac{1}{2},
\frac{1}{2},\dots, \frac{L_z - 1}{2}$, for a total of $L_z$ layers. The interface is induced by
walls of fixed spins $\sigma=+1$ at the top ($z = (L_z + 1)/2$) and $\sigma=-1$ at the bottom ($z =
-(L_z + 1)/2$) planes of the lattice; these boundary conditions energetically favour parallel
alignment of the interface with the $x\text{-}y$ plane, with the `+' phase in the upper half of the
volume, $z>0$. We focus on slab-like lattice geometries, with  $L_x, L_y \gg L_z$ and $L_x =
L_y\equiv L$, so that the system is confined between the two walls, and the scaling length scale for
the interfacial width is $L_z\equiv L_{\perp}$ \cite{LipowskyFisher,Whiskers}.

Time evolution of the system proceeds under Kawasaki \cite{Kawasaki} spin-exchange dynamics, which
conserve the magnetisation, or equivalently the number of lattice-gas particles, locally. Elementary
simulation moves consist of exchanging the values of two randomly chosen nearest-neighbour spins
with probability $p$; in the lattice gas, this corresponds to a particle moving to a neighbouring
empty lattice site. The key features of these dynamics are their conservation of order parameter,
and their locality. These attributes are desirable from the point of view of simulating particle
movement (albeit crudely) and the competition between diffusive motion and driven motion.
\begin{figure}[ht] 
\begin{center}
\includegraphics[scale=0.4]{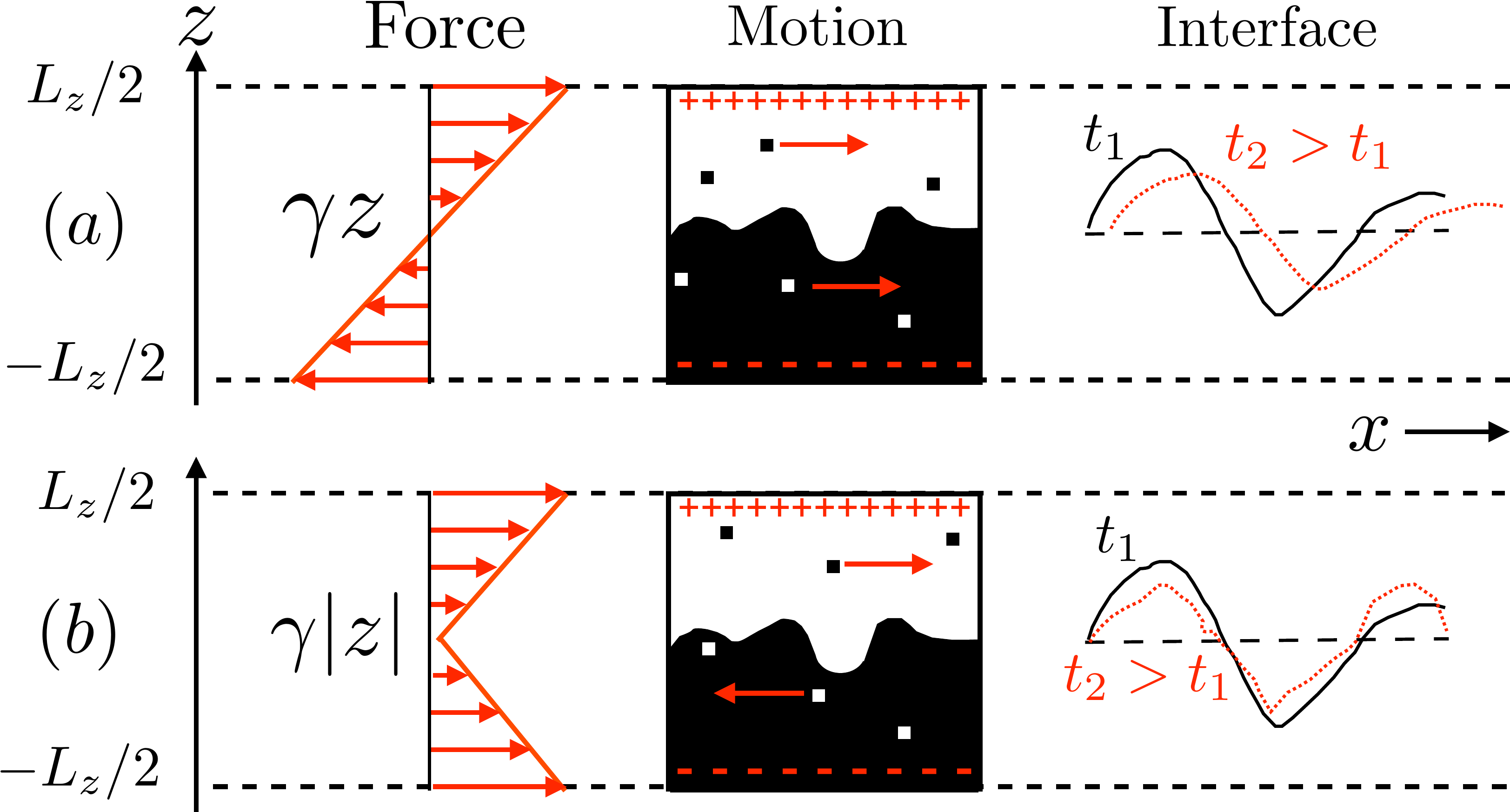}
\caption{(Color online) Illustration of the force field, microscopic particle motion, and
coarse-grained interface motion in the model system. (a) Shear-like driving field $F_x(z) = \gamma
z$. The `intruders' of one phase into the other move in the \emph{same} direction in both the upper
and lower halves of the system. As explained later, the interface displays transport. (b) V-shaped
driving field: the intruders now move in opposite directions. No interfacial motion occurs;
interfacial fluctuations decay and new ones appear as time passes.}
\label{fig:1}
\end{center}
\end{figure}
In the general case, an external force field $\mathbf{F}(z) = (F_x(z), F_y(z), 0)$ acts on the
system, driving in the $x\text{-}y$ plane. This field alters the Monte Carlo acceptance rates, to
produce a modified Metropolis rate,
\begin{equation}
\label{eq:rates}
p = \min \left\{ 1, \exp \left[- \beta (\Delta H + \Delta W) \right] \right\}.
\end{equation}
Here, $\beta = 1/k_{\rm B}T$ is the inverse temperature (the Boltzmann constant is set to unity),
and $\Delta H$ is the change in internal energy from the proposed exchange. $\Delta W$ is the work
done by or against the external force field; for $\Delta W = 0$, the above rate reduces to the
standard Metropolis one, which samples thermal equilibrium states. We are interested in the case of
non-zero $\Delta W$, when the system will reach a non-equilibrium steady state. The system is
immersed in a heat bath at constant temperature $T$, into which the work done is dissipated. The
driving field is related to the work term by
\begin{equation}
\label{eq:delta_w}
\Delta W = -J~ \boldsymbol\delta \cdot \mathbf{F}(z) (\sigma_i - \sigma_j) / 2,
\end{equation}
where $i = (x + \delta_x,y + \delta_y,z)$, $j = (x,y,z)$, and the displacement vector between spins
$i$ and $j$ is $\boldsymbol\delta = (\pm 1,0,0)$ or $(0,\pm 1,0)$. In the following we will consider
the forms of the driving field $\mathbf{F}$ which are  perhaps the most relevant experimentally. Our
main focus is the case of `shear-like' linear variation of driving field with $z$, and the field
acting in the $x$-direction only, such that the field components are $F_{x}(z) \equiv \gamma z$,
$F_{y}(z) \equiv 0$. Thus exchanges along $x$ are enhanced or suppressed, while exchanges in the $y$
and $z$ directions proceed with equilibrium rates ($\Delta W = 0$). We also study the case of a
V-shaped spatial dependence, $F_x(z) \equiv \gamma \left| z \right|$, $F_y(z) \equiv 0$, such that
the drive acts in the same direction throughout the system. Some results for a spatially uniform
driving field in the $x$ direction, $F_x \equiv f = \textrm{const}$, $F_y \equiv 0$, and a step-like
field, $F_x(z) = \textrm{sgn}(z)\cdot f$, are also included.

We have carried out extensive Monte Carlo (MC) simulations of the above model using single-spin and
multi-spin \cite{NewmanBarkema, GemmertBarkema} coding techniques, in the latter case generalizing
the driven multi-spin algorithm used previously to 3$d$ systems. The multi-spin method we have
adopted allows simulation of 64 independent systems (and hence greatly enhanced statistics) on a
64-bit computer system, using efficient bitwise operations. The state of an Ising spin may be
represented by one bit; thus the values of a particular site in 64 different systems can be stored
in a 64-bit variable (the natural word size). Bitwise operations operate on all bits of a variable
at once, and are computationally cheap instructions. By combining these operations appropriately,
and generating random bits with the required probabilities, the desired acceptance rates can be
produced.

Single-spin results provide a check as to the correctness of the multi-spin implementation.
Parallelization via domain decomposition was employed to speed up simulations; the lattice was
sub-divided along the $x$ direction, and appropriate synchronization used when exchanging spins on
and near the boundary between two domains. Data processing was also parallelized, owing to the large
quantity of data produced by the multi-spin algorithm. Results reported here are from total run
lengths of $2-4\times 10^7$~MC sweeps ($L_x \times L_y \times L_z$ trial moves), the slow evolution
of Kawasaki dynamics to a steady state proving to be less of a problem in $3d$ than in $2d$
\cite{LM7Proc}.

The majority of the results shown here are for a system size $L_x = L_y = 128$ and $L_z = 10$ or
$L_z = 20$, at a temperature $T/T_c = 0.75$, where $T_c \approx 4.5115$ ($\beta_c =0.2216544(3)$) is
the bulk critical temperature of the equilibrium 3$d$ Ising system \cite{IsingTc3d}. We have checked
that going to larger values of $L_x = L_y \leq 192$ has a minor effect on the results only for the
smallest considered  wall separation $L_z$, i.e., as long as the longitudinal correlation length
$\xi_{\parallel}$, which grows exponentially with $L_z$, is less than $L_x, L_y$. We have also
varied the temperature, firstly to investigate the effect of an increase to $T/T_c = 0.90$, and
secondly to study the behaviour near and below the roughening transition. For an equilibrium system
in the thermodynamic limit, the roughening temperature is $T_R \approx 2.454J$ \cite{HP}. In
a finite system, the (pseudo-)transition will occur at a shifted value of temperature, which will be
governed by the system size \cite{BarberFSS}. As temperature is increased in the smooth regime,
either the interfacial width will reach the scale of $L_z$, or the lateral correlation length will
reach $L_x$ or $L_y$ -- in either case, the interface reaches the rough regime. We have thus covered
a range of temperatures around the roughening temperature in the simulations, from $T/T_c = 0.4$ to
$T/T_c = 0.6$. The roughening transition belongs to the universality class of the
Kosterlitz-Thouless transition \cite{KT-trans}. The renormalization group method of Kosterlitz
\cite{Kosterlitz} showed that the correlation length $\xi_{\parallel}$ diverges very rapidly as
$T\to T_R$ from below:
\begin{equation}
\xi_{\parallel}=A\exp\left[\frac{B}{\sqrt{(T_R/T-1)}}\right],
\end{equation}
where $A$ and $B$ are non-universal parameters. The numerical values for these constants obtained
from MC simulation studies of the Ising interface in $3d$ are $A=0.80(1)$ and $B=1.01(1)$
\cite{HP}. The shift of the pseudo-roughening temperature can be estimated from the
condition $\xi_{\parallel}\simeq L_x(=L_y=L)$, which yields $(T_R-T_R(L,L,L_z))/T_R(L,L,L_z)\equiv
\Delta T \simeq \left( B/\ln (L/A) \right)^2$. This is a very weak dependence on $L$ and for our
system size it gives $\Delta T \simeq 0.04$.  At the same time the width of the interface diverges
upon approaching the  bulk roughening temperature, as $w^2\sim \ln (\xi_{\parallel})$. The condition
 $w \simeq L_z$  yields $\Delta T \simeq \left( B/(L_z^2-\ln A)\right)^2$, which is a much stronger
dependence on the finite dimension $L_z$ than that obtained from the condition involving
$\xi_{\parallel}$. For  $L_z=10$, $\Delta T \simeq 10^{-4}$, therefore we conclude that the shift of
the roughening transition is governed by $\xi_{\parallel}$. Using that estimate of $\Delta T$ gives
$T_R(L,L,L_z) / T_c \simeq 0.52$, so the range of simulation temperatures should include the equilibrium
pseudo-transition temperature.

%%%%%%%%%%%%%%%%%%%%%%%%%%%%%%%%%%%%					RESULTS  			%%%%%%%%%%%%%%%%%%%%%%%%%%%%%%%%%%%%%%%%
%%%%%%%%%%%%%%%%%%%%%%%%%%%%%%%%%%%%%%%%%%%%%%%%%%%%%%%%%%%%%%%%%%%%%%%%%%%%%%%%%%%%%%%%%%%%%%%%%%%%

\section{Results}
\subsection{Structure} \label{sec:static_results}

We first investigate the interfacial structure of the driven 3$d$ Ising model in order to see
whether also in three dimensions the effective action of drive is to increase the confinement of
the interface.

\subsubsection{Magnetisation profiles} \label{subsub:profile}

\begin{figure}[ht] 
\begin{center}
\includegraphics[scale=0.8]{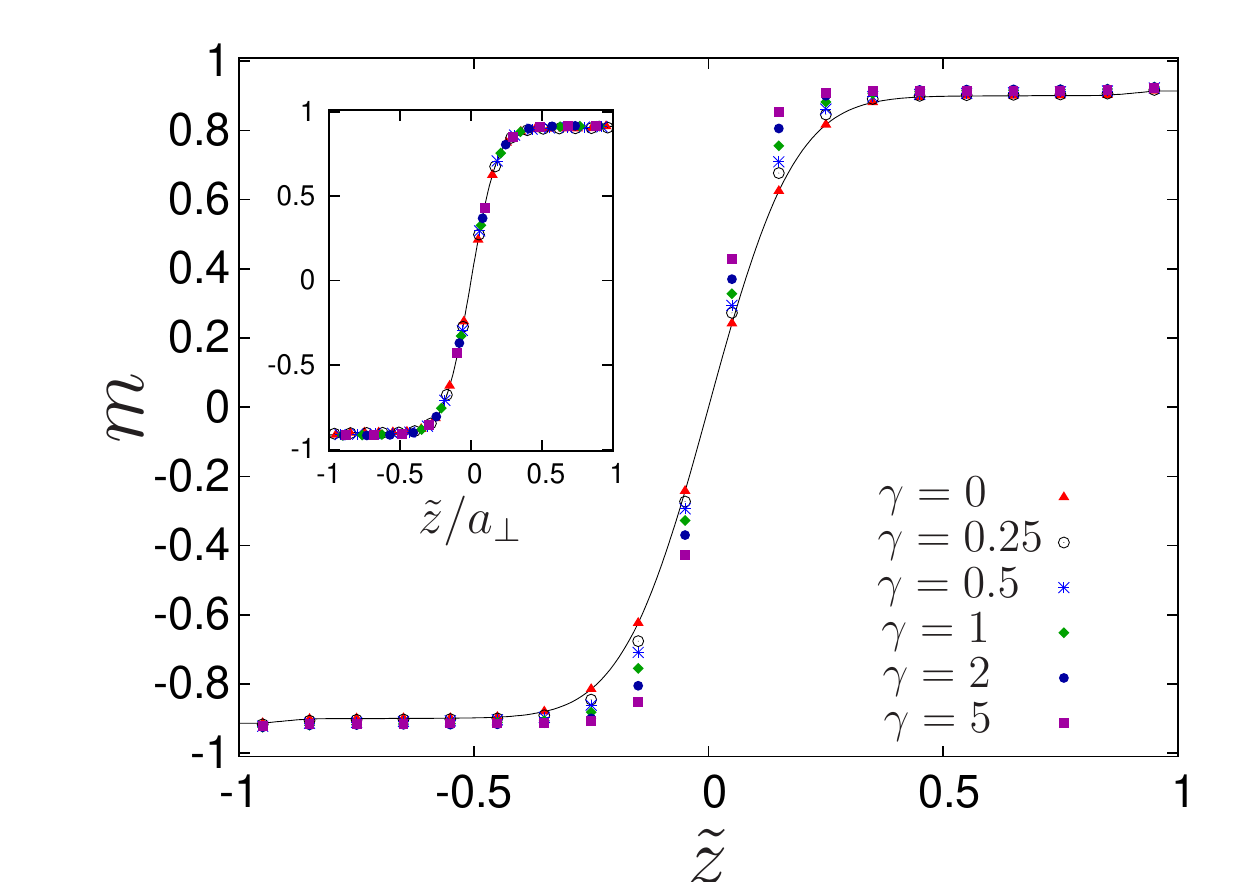}
\caption{(Color online) Magnetisation profiles $m(\tilde{z})$ as a function of a scaled coordinate
$\tilde{z}$ between the walls. The system is $L_x = L_y = 128, L_z = 20$, at a temperature $T/T_c =
0.75$. Results for equilibrium simulations ($\gamma=0$) with Kawasaki dynamics are shown (by symbols
and lines), as well as for shear-like drive with several values of $\gamma$ (with symbols only).
Results of rescaling to equilibrium are plotted in the inset; rescaling factors are $a_{\perp} =
0.88(5), 0.81(46), 0.71(8), 0.61(4), 0.51(28)$ for $\gamma = 0.25, 0.5, 1, 2, 5$, respectively.
Error bars are of order or smaller than the line thickness or symbol size, and are not shown.}
\label{FigMag}
\end{center}
\end{figure}

The magnetisation profile along the $z$ axis is calculated as
\begin{equation}
m(z) = \frac{1}{L_x L_y} \left< \sum_{x,y} \sigma(x,y,z) \right>
\end{equation}
where the angles denote an average in the steady state. In a phase separated system $m(z)$ changes
sign across the interface, and attains values close to $\pm 1$ near the upper and lower walls
respectively. In Fig.~\ref{FigMag} we plot the magnetisation as a function of the scaled variable
$\tilde{z} = 2z / L_z$. Upon applying either shear-like or V-shaped drive, the magnetisation profile
becomes `sharper': $m(\tilde{z})$ changes sign more rapidly in the interfacial region, and there is
a more extended flat region near either wall. The size of this effect increases with increasing
$\gamma$. These trends are the same as in two dimensions \cite{Shear2dPaper}, but for given driving
strength, we find that the magnitude of the effect is smaller in 3$d$. 

It is possible to rescale the driven profiles to collapse back onto the equilibrium result: see
Fig.~\ref{FigMag}. We interpret this as the drive acting to reduce the effective distance between
the walls from $L_z$ to $L^*_z$, and thus to effectively increase the confinement of the system.
This increase may be quantified by introducing the scaling $m(a_{\perp}\tilde{z}) \approx m_{\rm
eq}(\tilde{z})$, where $a_{\perp} = L^*_z / L_z$ is the ratio of the effective and actual wall
separations. For shear-like drive with $\gamma = 1.0$, we find $a_{\perp} = 0.71(8)$, whereas for
two dimensions \cite{Shear2dPaper}, for the same value of $\gamma$ ($\gamma$ is named $\omega$
there), $a^{2d}_{\perp} = 1/3.4 = 0.29$. One may more formally express this in terms of a finite
size scaling function for the profile. The scaling relation for an equilibrium fluid confined
between two walls was proposed by Fisher and de Gennes \cite{FdG}. In the absence of a bulk field it
reads:
\begin{equation}\label{eq:mscaling_0}
m(z,T,L_z)_{\rm eq}=m_{\rm b}(T)
{\cal {\tilde M}}_{\rm eq}\left(\frac{z}{\xi_b(T)},\frac{L_z}{\xi_b(T)}\right)=m_{\rm b}(T)
{\cal M}_{\rm eq}\left(\frac{z}{L_z},\frac{L_z}{\xi_b(T)}\right),
\end{equation}
where $\xi_b(T)$ is the bulk correlation length, and $m_{\rm b}(T)$ is the spontaneous magnetisation
in bulk. ${\cal {\tilde M}}_{\rm eq}$ and ${\cal M}_{\rm eq}$ are finite-size scaling functions:
${\cal M}_{\rm eq}(u,w)$ is obtained from ${\cal {\tilde M}}_{\rm eq}({\tilde u},w)$ by changing the
first scaling variable $\tilde{u} = uw$. Thus in equilibrium the shape of the scaling function can
be varied by changing the wall separation at fixed $T$ or, equivalently, by changing the temperature
at fixed $L_z$. We find  that driving changes the shape of the interfacial profile at fixed
temperature and $L_z$ such that
\begin{equation}\label{eq:mscaling}
\frac{m(z,T,L_z)}{m_{\rm b}(T)} \approx 
{\cal M}_{\rm eq}\left(\frac{z}{L_z^*},\frac{L_z*}{\xi_b(T)}\right) + 
{\cal M}_{\rm corr}(z) \textrm{\qquad with } L_z^* < L_z,
\end{equation}
where ${\cal M}_{\rm corr}$ is a boundary correction that decays away from the walls on the scale of
$\xi_b$. Relation (\ref{eq:mscaling_0}) admits (in the scaling regime) another interpretation of the
result (\ref{eq:mscaling}), namely, as the drive acting to reduce the effective temperature of the
system at fixed actual distance between the walls. The estimation of the rescaling
factor $a_{\perp}$ is obtained by rescaling the driven data to spline-interpolated equilibrium
curves and minimizing the associated chi-squared statistic -- the value of $a_{\perp}$ at the
minimum is the optimal value. For the $L_z = 20$ system, the 8 points in the
centre of the system were included in the rescaling procedure -- points further out are subject to
stronger wall interaction effects.
Both ways of driving yield very similar rescaling factors; more pronounced differences are observed for
smaller values of $\gamma$ and lower temperatures with the conclusion that effective confinement is
stronger for the V-shaped drive. Rescaling fails for lower temperatures  closer to and below the 
equilibrium bulk roughening transition temperature.

\begin{figure}[ht] 
\begin{center}
\includegraphics{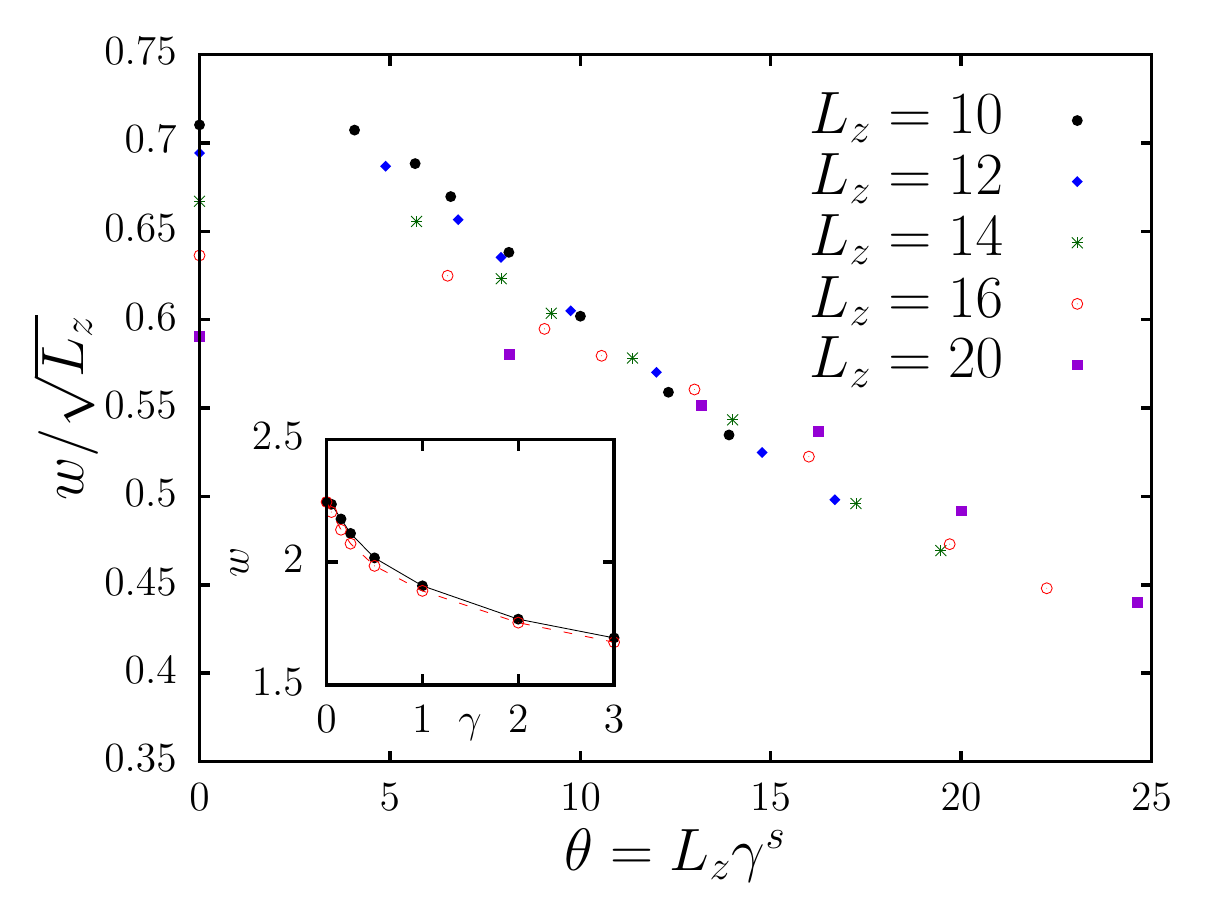}
\caption{(Color online) Scaling behaviour of the interfacial width $w$ at a temperature
$T/T_c=0.75$, as a function of the scaling variable $\theta = L_z\gamma^s$, with $s=0.3$. The width
is scaled by $\sqrt{L_z}$ according to the equilibrium relation $w / \sqrt{L_z} = {\rm const.}$, as
discussed in the text. Different point types correspond to differing values of $L_z$, from 10 to 20,
as indicated. Inset: variation of the width with drive gradient $\gamma$, for shear-like drive
(filled circles) and V-shaped drive (open circles).}
\label{FigWidth}
\end{center}
\end{figure}

The magnetisation profile may also been used to study the behaviour of the interfacial width. We
measure the width via the second moment of $dm / dz$ and study its variation with driving strength,
wall separation $L_z$, and temperature. Upon increasing driving strength $\gamma$ or $f$, the width
reduces, as expected from the results for the full profile. For shear-like drive, we are also able
to obtain data collapse for the behaviour of $w / \sqrt{L_z}$ as a function of a scaling variable
$\theta = L_z \gamma^s$. Here $s$ is an adjustable exponent. The division of the width by
$\sqrt{L_z}$ corresponds to the expected equilibrium behaviour \cite{LipowskyFisher, Whiskers,KKB},
so that for $\theta \to 0$, $w / \sqrt{L_z} \to {\rm constant}$. The scaling behaviour of the width
is shown in Fig.~\ref{FigWidth}, for fixed temperature $T/T_c = 0.75$, and a variety of wall
separations and drive gradients in the ranges $10 \leq L_z \leq 20$, $0 < \gamma \leq 2$. Previously
we obtained data collapse as a function of the same scaling variable in the 2$d$ system
\cite{Shear2dPaper} (there the width was scaled by the $2d$ equilibrium behaviour, $w \sim L_z$).
Remarkably, we obtain data collapse for the \emph{same} value of exponent $s = 0.3$ as for the $2d$
system. From Fig.~\ref{FigWidth} we see that for small $\gamma$ at the larger values of $L_z = 16$
and 20, the data collapse is lost -- we believe that this is because one moves out of the confined
regime with $w \sim \sqrt{L_z}$ for these parameters. For these wall separations, the longitudinal
correlation length $\xi_{\parallel}$ becomes comparable to the linear dimension $L$ of the
interface, and the system crosses over to the regime where the dominant length scale is $L$. This
does not require a large increase in $L_z$, because $\xi_{\parallel} \simeq \exp(\kappa L_z / 4)$,
where the transverse length scale $\kappa^{-1} = \xi_{\rm b}$ for Gaussian interface fluctuations
\cite{Fisher, KKB}. Data collapse is regained for larger values of $\gamma$, because the effective
wall separation $L_z^* < L_z$ is the controlling length scale (from the discussion of the
magnetisation profile above), and $L_z^*$ is small enough for the system to be in the ``confined
regime''. The inset of Fig.~\ref{FigWidth} shows the variation of the width with drive gradient
$\gamma$ for shear-like and V-shaped drive. The trends are rather similar, with the width for given
$\gamma$ very slightly smaller for V-shaped drive -- this is consistent with the previous conclusion
that the confinement is stronger for this drive type.

\subsubsection{Spin-spin correlation functions} \label{subsub:spinspin}

\begin{figure}[ht]
\centering
\subfigure{
\includegraphics[scale=0.8]{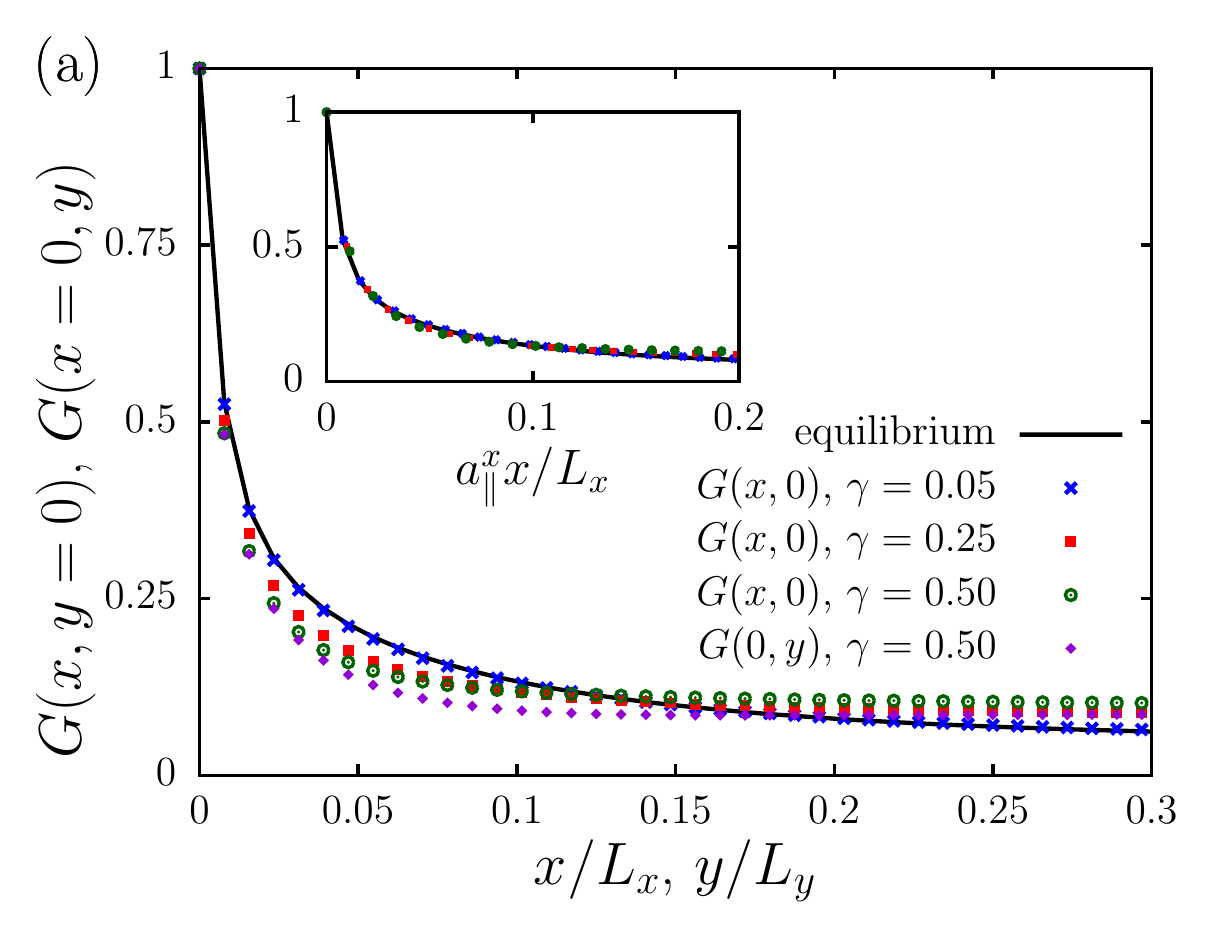}
\label{FigSpinspin_rescaling}
}
\subfigure{
\includegraphics[scale=0.8]{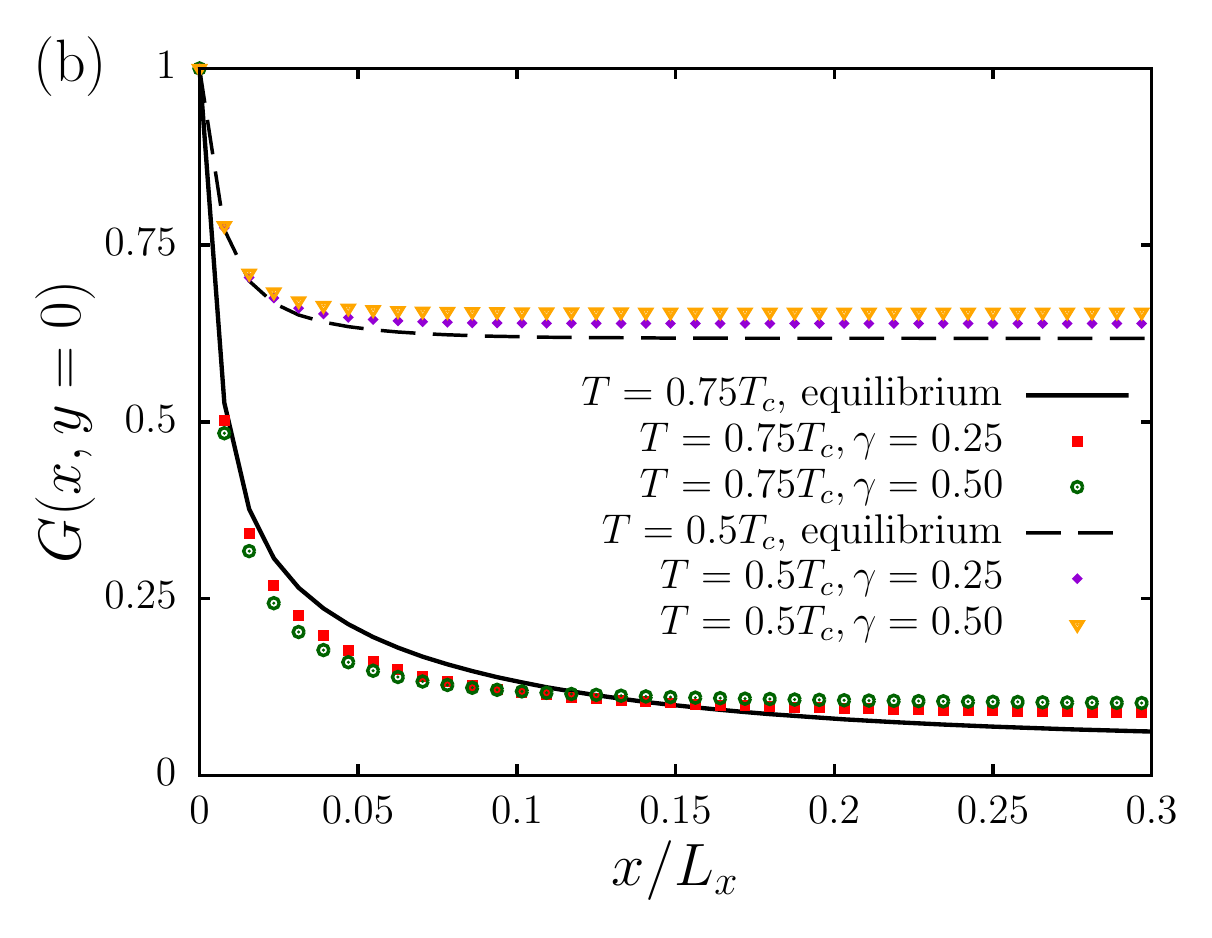}
\label{FigSpinspin_temp}
}
\caption{
(Color online) Spin-spin correlation functions $G(x,y=0)$ and $G(x=0,y)$ as a function of scaled
separation $x/L_x$ or $y/L_y$. (a) Results for a $128 \times 128 \times 10$ system at $T/T_c =
0.75$. The equilibrium result is shown, as well as driven results for shear-like drive at  several
values of drive gradient $\gamma$, as indicated. In the inset the driven results for $G(x,y=0)$ are
rescaled via the parameter $a_{\parallel}^{x}$, as described in the text; rescaling factors are
$a_{\parallel}^x$ = 0.95, 0.79, 0.69 for $\gamma = 0.05, 0.25, 0.5$, respectively. The maximum
separation for the correlation functions is $x/L_x = y/L_y = 0.5$, due to the periodic boundary
conditions; beyond the displayed region the functions have essentially reached their asymptotic
values. (b) Results for $G(x,y=0)$ for $128 \times 128 \times 10$ systems at $T/T_c = 0.75$ and
$T/T_c = 0.5$, for equilibrium and two strengths of shear-like drive. Below the equilibrium
roughening temperature, i.e., for $T/T_c = 0.5$, the asymptotic value is large, and increases with
stronger drive.}
\label{FigIface}
\end{figure}

In order to characterize the behaviour of the sytem on the two-body level, we first consider the
microscopic interface pair correlation function, i.e., the spatial spin-spin correlation function
at the midplane, defined as
\begin{equation} \label{EqGCorr}
G(x, y, z = L_z/2) = \frac{1}{L_xL_y}
	\left< \sum_{x', y'} \sigma(x',y', L_z / 2)\sigma(x'+x,y'+y, L_z / 2) \right>,
\end{equation}
which depends on separations in both the $x$ and $y$ directions. Comparing the specific cases
$G(x,y=0)$ and $G(x=0,y)$ provides information on the anisotropy in the $x$ and $y$ directions. In
equilibrium, $G(x,0) = G(0,y)$ for $L_x = L_y$, but for driven systems, the two functions may
differ. Results in Fig.~\ref{FigIface}a show that shear-like drive causes \emph{both} $G(x,0)$ and
$G(0,y)$ to decay more quickly and for larger separations to saturate at larger asymptotic values
than in equilibrium. In the $x$ direction, this finding is in agreement with hydrodynamics results
\cite{ThieBick}; however in that study, the correlation length in the $y$ direction was found to
\emph{increase} under shear, contrary to the trend in our system. We defer further exploration of
this difference to the discussion of the height correlations below, since the height variable
provides a more direct point of comparison between the systems.

As in the 2$d$ case, the spin correlation functions at intermediate separations may be transformed
to the equilibrium result via a rescaling of the lateral coordinate, $x$ or $y$: $G(a_{\parallel}^{
x}x, 0) \approx G^{\rm eq}(x, 0)$ and $G(0, a_{\parallel}^{y}y) \approx G^{\rm eq}(0, y)$
\cite{Shear2dPaper}; see the inset of Fig.~\ref{FigIface}a for rescaling results for $G(x,0)$. The
rescaling factors are obtained via the same method as for the magnetisation profile; in this case,
very small values of $x$ or $y$ are cut-off in the procedure, as are the tails of the functions, so
that the rescaling procedure is carried out over $2 \leq x,y \leq 16$. The $a_{\parallel}$
parameters may be interpreted as ratios of lateral interfacial correlation lengths in and out of
equilibrium:
$a_{\parallel}^{x} = \xi_{\parallel}^{ x} / \xi_{\parallel}^{x, \rm eq}$.
Rescaling the driven results produces $a_{\parallel}^{x} < 1$ and $a_{\parallel}^{y} < 1$, so that
the correlation length is reduced under drive -- this tallies with the faster decay evident from
Fig.\ \ref{FigIface}a, which shows results for equilibrium and shear-like drive. The $x\text{-}y$
anisotropy may be measured by the ratio $a_{\parallel}^{y} / a_{\parallel}^{x}$; this is
consistently slightly smaller than unity, leading to the surprising conclusion that the correlations
are slightly \emph{more} suppressed in the $y$ (vorticity) direction. As with the magnetisation
profile, the effect of drive is much weaker in three dimensions than in two -- for example, in 2$d$,
$a_{\parallel}^{x} = 1/1.27$ for $\gamma = 0.025$, while in 3$d$, $a_{\parallel}^{x} = 1/1.271
\approx 0.79$ for $\gamma = 0.25$: a ten times larger field gradient is required to produce a
comparable confinement. As a result, the rescaling procedure works for much larger values of
$\gamma$ than in 2$d$ -- a stronger drive is required to push the system into a regime which is too
far from equilibrium for rescaling to be possible. The V-shaped drive has a similar effect on the
interfacial correlations: for example, with $\gamma = 0.25$, $a_{\parallel}^{x} = 0.74$,
corresponding to a slightly stronger confinement effect than with shear, consistent with the
findings for the magnetisation profile.

Fig.~\ref{FigIface}b shows the effect of varying the temperature on the interfacial correlations.
Lowering the temperature to $T/T_c = 0.5$ (below the roughening transition in a bulk equilibrium
system) in equilibrium results in correlations $G(x, 0)$ that decay only  to $\sim 0.6$ for the
largest separations. Since on the lattice the average interface position lies \emph{between} two
lattice points (for zero overall magnetisation), one measures the correlations just either side of
the interface (which side does not matter, due to symmetry). Thus at zero temperature,
$G(x,y,z=L_z/2) = 1$ for all $x,y$, since the interface is perfectly flat at $T = 0$. This explains
the observed increase of asymptotic values of $G(x, 0)$ for low $T$. Moreover, the width of smooth
interfaces is of order of the bulk correlation length, which at low temperatures is $\sim$ 2-3
lattice spacings. Therefore, for low temperatures $G(x, 0)$ essentially measures
correlations in the bulk-like phase. We also see from Fig.~\ref{FigIface}b that driving the system
enhances the asymptotic value further for $T/T_c = 0.5$, which indicates that at fixed temperature the
bulk-like phase is more ordered under drive than in equilibrium. As for the magnetisation profile,
rescaling does not work for the low temperatures -- the drive affects the asymptotic value more
than the decay rate for these temperatures.

\subsubsection{Height-height correlation functions} \label{subsub:height}

\begin{figure}[ht]
\begin{center}
\includegraphics{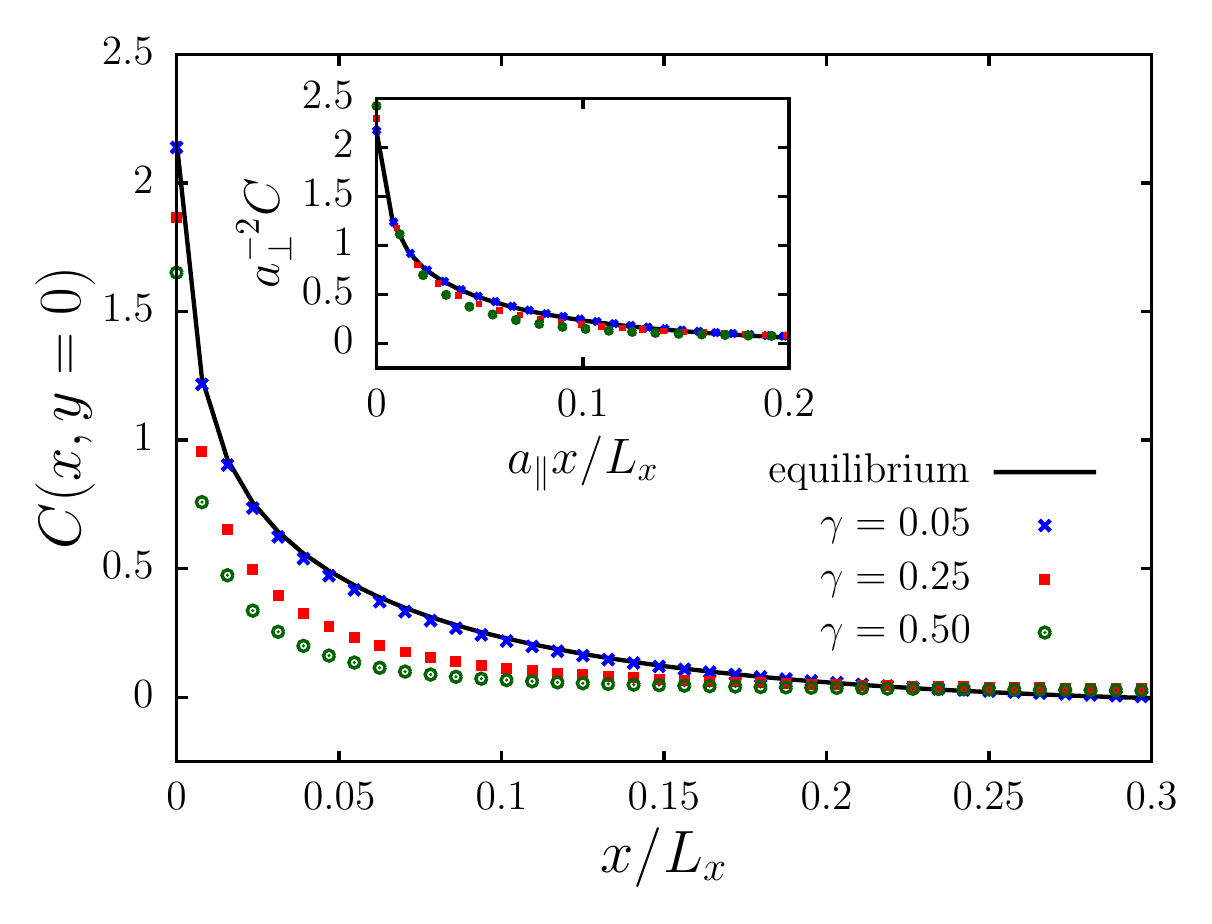}
\caption{(Color online) Height-height correlation function $C(x,y=0)$ as a function of separation
$x$, for a $128 \times 128 \times 10$ system at $T/T_c = 0.75$. In the inset, data for non-zero
drive are rescaled to the equilibrium result via the relation $a_{\perp}^{-2}C(a_{\parallel}^{x}x)
\approx C_{\rm eq}(x)$ given in the text, where the values of $a_{\perp}$ and $a_{\parallel}^{x}$
are obtained from the rescaling of the magnetisation profile and spin-spin correlation functions,
respectively.}
\label{FigHeight}
\end{center}
\end{figure}

We now turn to the interfacial height-height pair correlation function. The interface height is
obtained via a coarse-graining procedure, which produces a single-valued height function $h(x,y)$
from the real microscopic configuration. The latter contains `bubbles' of one phase in the other,
and `overhangs' at the interface, meaning a height cannot be defined from it directly. To
coarse-grain, we use a simple method which we found to be successful in the 2$d$ system, where it
gives results equivalent \cite{Shear2dPaper} to a more complicated coarse-graining method
\cite{DeVirgiliis}. The height $h(x,y,t)~=~(1/2)\sum_{z}\sigma(x,y,z,t)$ is simply a sum of the spins
over a column. Thus when there are equal numbers of `+' and `-' spins in a column, $h(x,y)=0$, while
when there is a majority of one species, $h \neq 0$. The general height-height correlation function
depends on spatial separations $x$, $y$ and on temporal displacement $t$:
\begin{equation}
\label{eq:cxyt}
C(x,y,t) = \frac{1}{L_xL_y} \left< \sum_{x', y'} h(x',y', t') h(x'+x,y'+y, t'+t) \right>,
\end{equation}
where the angles indicate an average over time. We first consider the equal-time correlations, with
one of the spatial separations set to zero: Fig.\ \ref{FigHeight} shows results for $C(x,y=0,t=0)$.
In two dimensions, $C(x,t=0)$ (also $G(x)$) in equilibrium exhibits strong anti-correlated regions
for medium-to-large separations, presumed to be finite size effects \cite{LM7Proc}. These are not
present in $3d$ for the system sizes considered -- the functions decay to zero without becoming
signifcantly negative, indicating less severe finite size effects; an explanation may be the
following. With conservative dynamics, a positive-height `bump' must be accompanied by one with
negative height, since $\sum h \equiv 0$. In $2d$, these must lie on the same $x\text{-}z$ layer
(the only one), so an anti-correlation is measured in $C(x)$. However in $3d$ there are $L_y$
$x\text{-}z$ layers, so the pair may be located in different layers, meaning $C(x,y=0)$ does not
necessarily display anti-correlations. Turning to the driven cases, we see that applying shear-like
drive leads to more rapid decay of $C(x,0,0)$, as well as a smaller initial value $C(x=0$) (a
measure of the interfacial width). The magnitude of this effect increases with increasing $\gamma$.
In this section, the results shown are for shear-like drive, but the conclusions also apply to
V-shaped drive -- as for other static quantities, the effect is similar to shear, with slightly
greater correlation-suppression.

As for other quantities, rescaling to the equilibrium result is possible in the same manner as in
$2d$. For the height correlations, the rescaling takes the form $a_{\perp}^{-2}C(a_{\parallel}^{x}x)
\approx C_{\rm eq}(x)$. The values of $a_{\perp}$ and $a_{\parallel}$ are those obtained from the
rescaling  of the magnetisation profile and the spin-spin correlation function for given simulation
parameters. In $2d$ this procedure was motivated by Weeks scaling in equilibrium
\cite{BedeauxWeeks}: $C_{\rm eq}(x)\approx w^2{\cal C}(x/\xi_{\parallel}^{eq})$, where $\cal{C}$ is
the scaling function, and since $w\propto L_z$ in $2d$, the correct rescaling involves $a_{\perp}^2$.
Although Weeks scaling does not hold in $3d$, this procedure works reasonably well for $\gamma
\lesssim 0.5$, except at zero separation, where the rescaled values become larger than the
equilibrium width. As with the spin correlations, this range of validity is much greater than in two
dimensions.

\begin{figure}[ht]
\begin{center}
\includegraphics{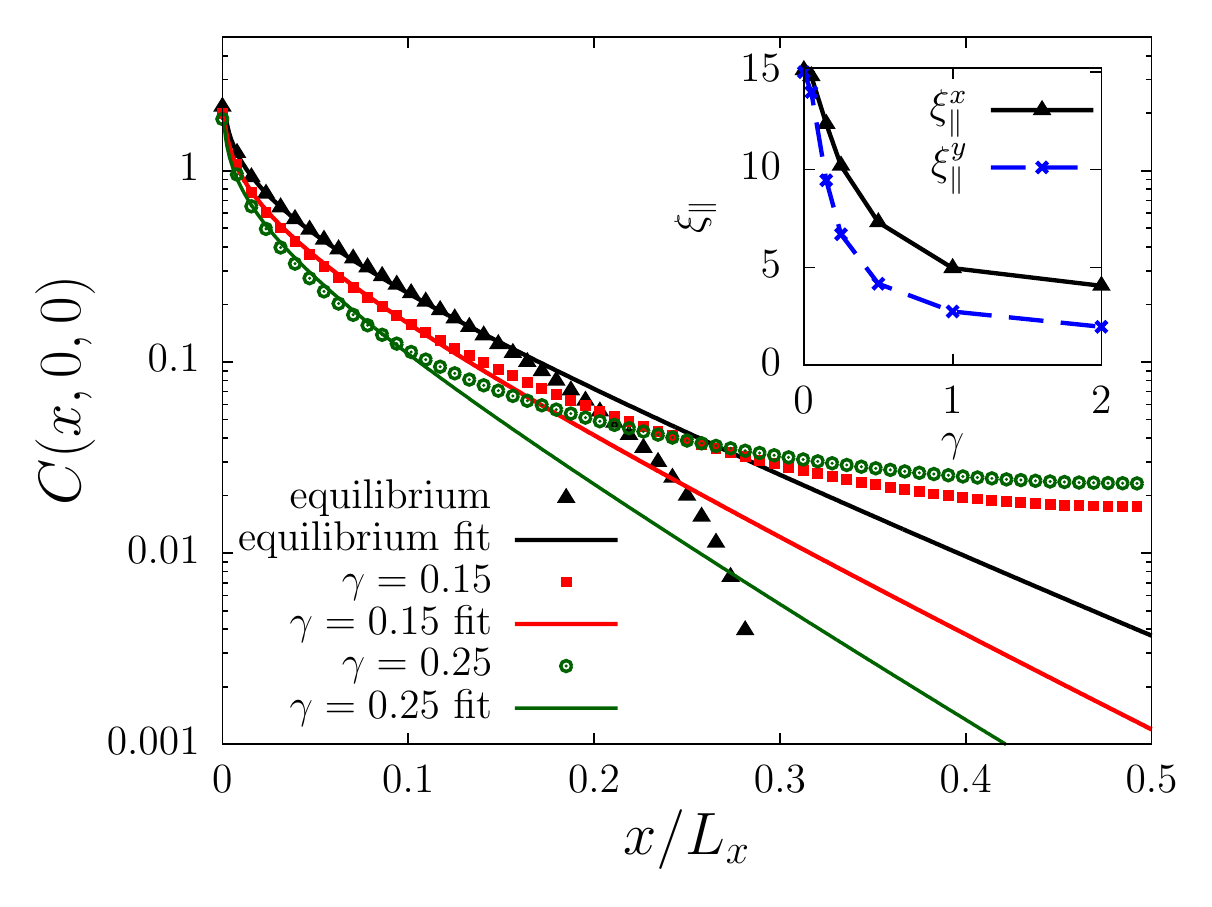}
\caption{(Color online) Fits of the equilibrium and non-equilibrium height correlation data for
$C(x,y=0,t=0)$ to the asymptotic capillary wave prediction given in the text. The system parameters
are the same as in Fig.~\ref{FigHeight}. Inset: correlation lengths $\xi_{\parallel}^{x}$
and  $\xi_{\parallel}^{y}$ (via fitting $C(x=0,y,t=0)$ data) obtained from the fits, as a function
of shear-like drive gradient $\gamma$.}
\label{FigBessel}
\end{center}
\end{figure}

Furthermore, we are able to fit the results for $C(x,0,0)$ and $C(0,y,0)$ for small-to-intermediate
separations to the \emph{equilibrium} capillary wave result for the height correlation function in
$3d$; see Fig.~\ref{FigBessel} for results for shear-like drive. This procedure was used previously
by Derks et \emph{al.} to describe their experimental data, where an excellent fit was obtained
\cite{DerksShear}. By integrating Eqn.~\eqref{eq:hcorr}, the (equilibrium) capillary wave result for
$C(x,0,0)$ in $d=3$ and in the limit  $L_x = L_y \to \infty$ is \cite{BedeauxWeeks}
\begin{equation}
C(x,0,0) = 
\frac{k_{\rm B}T}{2\pi\Gamma}K_{0} \left( \sqrt{\left(x / \xi_{\parallel}^{x} \right)^2 + \lambda^2} \right),
\label{eq:cx_cwt}
\end{equation}
where $K_0$ is the modified Bessel function of the second kind. The upper wave-number cutoff has
been sent to infinity in order to obtain an analytic result; in order to regularize the integral at
$x=0$, a shift $\lambda$ is introduced. Bedeaux and Weeks \cite{BedeauxWeeks} give $\lambda \approx
1 / (q_{\rm max} \xi_{||})$, where $q_{\rm max}$ is the original wave-number cutoff in the integral
\eqref{eq:hcorr}. Combining \eqref{eq:cx_cwt} with the capillary-wave result for the interfacial
width,
\begin{equation}
w^2 \equiv C(0,0,0) = \frac{k_{\rm B}T}{4 \pi \Gamma} \ln \left[1 + q_{\rm max}^2 \xi_{||}^2 \right],
\label{eq:capwave_width}
\end{equation}
we are able to substitute for the (unknown) $q_{\rm max}$ in terms of the width, and obtain a
fitting form with two parameters: the correlation length $\xi_{\parallel}$ and the pre-factor
$k_{\rm B}T / \Gamma$. In the above, we have specialised to separations in $x$ rather than the
radial distance $r$ usual in CWT, since isotropy is broken in the non-equilibrium situation. For
separations in $y$, the form for $C(0,y,0)$ is the same, but different values of the parameters are
expected -- i.e., the correlation length ($\xi_{\parallel}^{y}$) will be different, as will the
pre-factor. The intepretation of the latter quantity is difficult. Indeed, the interface tension is
an \emph{equilibrium} concept and cannot be carried over directly to non-equilibrium situations, so
the meaning of the pre-factor is not initially clear -- here we just note its anisotropy.

The equilibrium fit wanders off the data for larger separations; this may be due to a (less serious)
manifestation of the finite size effects encountered in $2d$, which were mentioned above, and the
conserved order parameter. (This is most obvious for the equilibrium data on the log scale in
Fig.~\ref{FigBessel}, where the data diverge as they approach zero and become negative). For the
driven cases, as drive becomes stronger, the fit works for a smaller range of separations -- the
example of shear-like drive is given in Fig.~\ref{FigBessel}. This trend is expected from the
findings for the rescaling procedures applied above -- initially the system is ``close enough'' to
equilibrium for CWT to be approximately applicable, but as drive increases, this ceases to be true.
From the fits we obtain the equilibrium and non-equilibrium correlation lengths in the $x$ and $y$
directions, $\xi_{\parallel}^{x}$ and $\xi_{\parallel}^{y}$-- see the inset of Fig.~\ref{FigBessel}
for their variation with $\gamma$. The trend of decreasing correlation length with increasing drive
strength mirrors the one found in $2d$, although there we were not able to obtain the correlation
length reliably, due to the difficulty of fitting the correlation function data over a reasonable
range. We also note that $\xi_{\parallel}^{y}$ is consistently smaller than $\xi_{\parallel}^{x}$,
in agreement with the earlier conclusions, based on the behaviour of the spin-spin correlation
functions, that correlations are slightly more strongly suppressed in the vorticity direction than
in the driving direction.

The suppression of correlations we find in the drive ($x$) direction is in agreement
with the hydrodynamics work of Thi{\'e}baud and Bickel \cite{ThieBick}, who studied phase separated
fluids between two walls under shear. This trend is also the same as in the 2$d$ Ising system, and
both microscopic ($G(x)$, spin-spin) and coarse-grained ($C(x)$, height) measures of correlations
give the same conclusion. Both the theoretical and simulation findings disagree, however, with the
experimental results of Derks et \emph{al.} \cite{DerksShear}, who found an \emph{increase} of
correlation length in the flow direction when shear was applied to a phase-separated colloid-polymer
mixture. The fact that we have used the same method of fitting the height correlation data to the
equilibrium capillary wave form as Ref.~\cite{DerksShear}, makes the method of comparison the same,
at least. 

Finally, we consider the pre-factor resulting from the fit of the height correlation data to the CWT
form \eqref{eq:cx_cwt}. In equilibrium, the pre-factor is proportional to $k_{B}T / \Gamma$, where
$\Gamma$ is the surface stiffness, which for a continuum fluid is the interfacial tension -- the
free energy associated with the interface. Out of equilibrium, this quantity is not defined, so the
meaning of the pre-factor resulting from the fit is not clear. Numerically, we find that the
pre-factor from fitting $C(x,0,0)$ is a decreasing function of drive gradient $\gamma$ for
shear-like drive. If one \emph{defines} a ``non-equilibrium surface tension'' from the CWT fit, and
further takes the temperature $T$ to be fixed (i.e., the value of the parameter in simulations), the
conclusion is then that this ``tension'' increases as the system is more strongly driven. This
procedure of defining an effective non-equilibrium surface tension via the CWT fit was the approach
taken in the analysis of the experiments of Derks \emph{et al.} \cite{DerksShear}, who also found
this tension to be an increasing function of shear rate. The CWT fits of experimental data in
Ref.~\cite{DerksShear} show that an increase in a ``non-equilibrium surface tension'' is accompanied
by an \emph{increase} in correlation length, but our simulations show the opposite relationship -- a
decreasing correlation length as the effective interfacial tension increases. Our findings seem to
be inconsistent with the CWT  result (see Eqn.~\eqref{eq:hcorr} and below) $\xi_{\parallel} \propto
\sqrt{\Gamma}$. However, in our system at equilibrium $\xi_{\parallel}\propto \sqrt{\Gamma}
\exp(L_z/(4\xi_b))$, so that the effective increase of the confinement due to driving (reduction of
$L_z$) wins over the effective increase of $\Gamma$.

\subsubsection{Structure factor} 
\label{subsub:4}

\begin{figure}[ht]
\begin{center}
\includegraphics{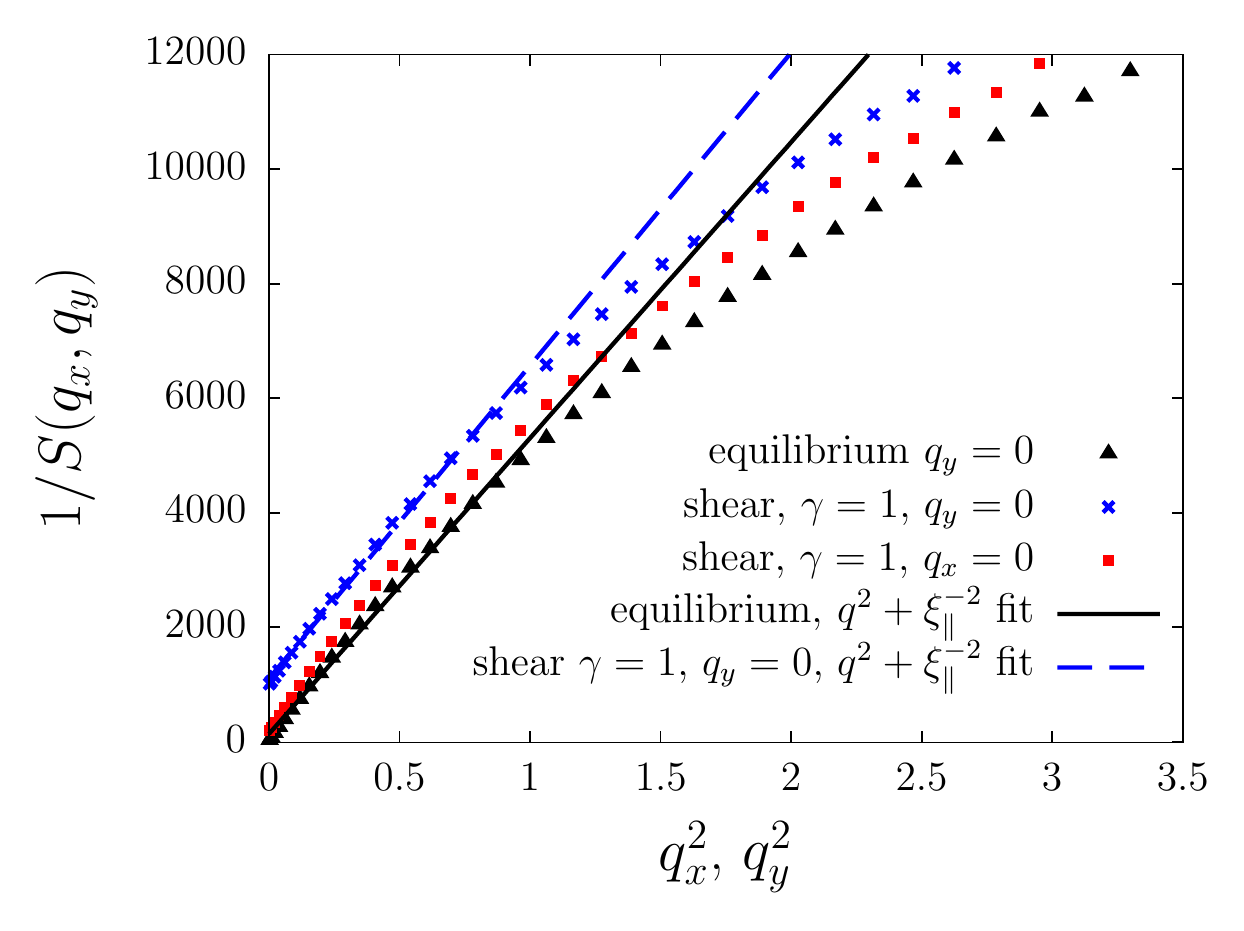}
\caption{(Color online) Structure factor $S(\mathbf{q})$, as defined in \eqref{eq:structfact}. Data
are shown for $S(q_x, q_y = 0)$ and $S(q_x = 0, q_y)$, as a function of the squared norm of the wave
vector, $q^2 = q_x^2$ or $q_y^2$, respectively. As before, the system dimensions are $128 \times 128
\times 10$, and the temperature is $T/T_c = 0.75$. Equilibrium data are shown, as well as for
shear-like drive with $\gamma = 1$, and fits to the CWT form are displayed for equilibrium and for
$S(q_x,0)$ with $\gamma = 1$, up to $q_x^2 = 1$.}
\label{FigStruct}
\end{center}
\end{figure}

To further complicate the situation, our finding of a decrease of correlation length in the
vorticity ($y$) direction is in disagreement with Ref.~\cite{ThieBick}, where an \emph{increase} was
found. Intruigingly, Thi{\'e}baud and Bickel found the structure factor $S(\mathbf{q}) = S(q_x,
q_y)$ to be \emph{unaffected} in the $q_y$ direction by the application of shear \cite{ThieBick}.
The same was concluded for the uniformly driven system from the analytic approach based on the
time-dependent Landau-Ginzburg functional by Leung \cite{Leung88}. The static structure factor in
our system is accessible via a two-dimensional spatial Fourier transform of the equal-time height
correlations,
$C(x,y,t=0)$:
\begin{equation}
S(\mathbf{q}) = \left| {\cal F} \left\{ C(x,y,t=0) \right\} \right|^2,
\label{eq:structfact}
\end{equation}
where ${\cal F}\left\{ \right\}$ denotes a two-dimensional spatial Fourier transform,
$\mathbf{q}~=~(2n_x\pi / L_x, 2n_y\pi / L_y)$, and $n_{x,y} = 0,1 \dots ((L_{x,y} / 2) - 1)$, so
that $q_x$ and $q_y$ lie on the range $0\dots (\pi - (2\pi / L_{x,y}))$. In Fig.~\ref{FigStruct} we plot
$1/S(\mathbf{q})$ for equilibrium and driven systems, along either the $q_x$ or $q_y$ direction, as
a function of $q^2 = |\mathbf{q}|^2$. From Eqn.~\eqref{eq:hcorr}, in equilibrium, one expects
$1/S(q) \propto \left(\Gamma / k_{\rm B}T\right) \left(q^2 +~ \xi_{\parallel}^{-2}\right)$. 
In Fig.~\ref{FigStruct} we fit the equilibrium data to this form. The data shown are along the
direction with $q_y = 0$, although we have checked that the equilibrium structure factor behaves the
same along the $q_y$ direction, as expected. For $q_x \lesssim 1$, this behaviour is indeed
observable in the simulations, but for $q_x \gtrsim 1$, the data diverges from the CWT prediction,
when other powers of $q_x$ presumably become important.

Turning to the non-equilbrium behaviour, we see that for shear-like drive, $S(\mathbf{q})$ is
affected (suppressed) in both the drive ($x$, blue crosses) \emph{and} vorticity ($y$, red squares)
directions, but the effect is smaller in the vorticity direction. These results are in disagreement
with the hydrodynamics results \cite{ThieBick}; however, since the effect in the drive direction is
stronger than in the vorticity direction, the latter effect could possibly be of higher order than
was considered in Ref.~\cite{ThieBick}. The data for shear-like drive along $q_y = 0$ are also fit
to the equilibrium CWT form in Fig.~\ref{FigStruct}; we see that as for equilibrium, the fit is
reasonable for $q_x \lesssim 1$. Additionally, the intercept at $q_x = 0$ is greater, indicating a
smaller lateral correlation length, as found in real space above. Indeed, one can compare the
parameters resulting from the CWT fits in real space and Fourier space. We find that the qualitative
trend for the correlation length is the same, but do not obtain quantitative agreement -- the values
obtained from the real space fits are consistently larger. These differences are expected -- for
equilibrium, they can be caused by the finite system size and lattice discretization effects. For
non-zero drive, the effect of deviations from CWT can be different in real and Fourier space.
Additionally, the fits in Fourier space are for small $q$ (long wavelengths), while the real-space
fits are for small separations, so the length scales the fits are applicable to is not necessarily
the same. For V-shaped drive, we find that for given drive gradient $\gamma$, the results are
similar to those for shear-like drive, with slightly greater suppression of the structure factor at
small $q$.

\subsection{Capillary wave transport} \label{sec:dynamic_results}

We now consider the dynamics of the height-correlation function defined in \eqref{eq:cxyt}. The
primary limit of interest is $C(x,y=0,t)$ which we find shows evidence of capillary wave transport
along the driving direction, for suitable forms of the current profile. Previously (in the context
of $d=2$) \cite{MovPaper}, we conjectured that the capillary wave fluctuations on an interface will
be transported by an external driving field, provided that \emph{the lateral order parameter current
has a component which is an odd function of distance from the interface}. Thus only \emph{purely
even} current profiles are expected to give no transport.

\subsubsection{Order parameter current} \label{subsec:dr1}

In the $3d$ system, the order parameter current profile is defined as $\mathbf{j}(z) =
\mathbf{j}_+(z) - \mathbf{j}_-(z)$, where $\mathbf{j}_{\sigma}(z)$ is the current profile of the
$\pm 1$ spin species. The components $j_{\sigma}^x(z)$ and $j_{\sigma}^y(z)$ of
$\mathbf{j}_{\sigma}(z)$ are the net number of spins of that species moving in positive $x$ and $y$
directions per unit time at perpendicular coordinate $z$. We also note that the Ising symmetry means
that the order parameter current can also be written as $\mathbf{j}(z) = 2\mathbf{j}_+(z)$; the
first definition may be applied to systems lacking the Ising symmetry, i.e., liquid-gas or
liquid-liquid interfaces.

\begin{figure}[ht]
\centering
\subfigure{
\includegraphics{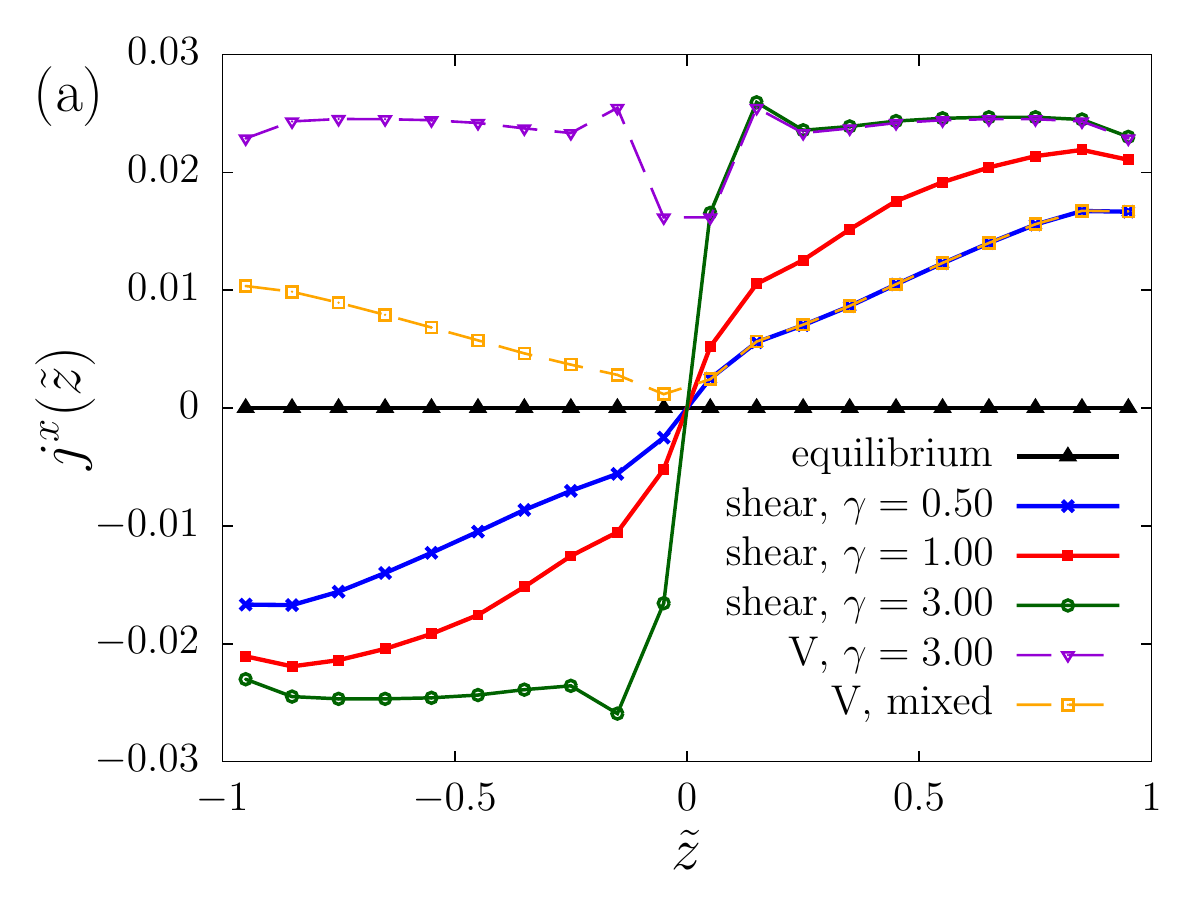}
\label{FigCurrent_075}
}
\subfigure{
\includegraphics{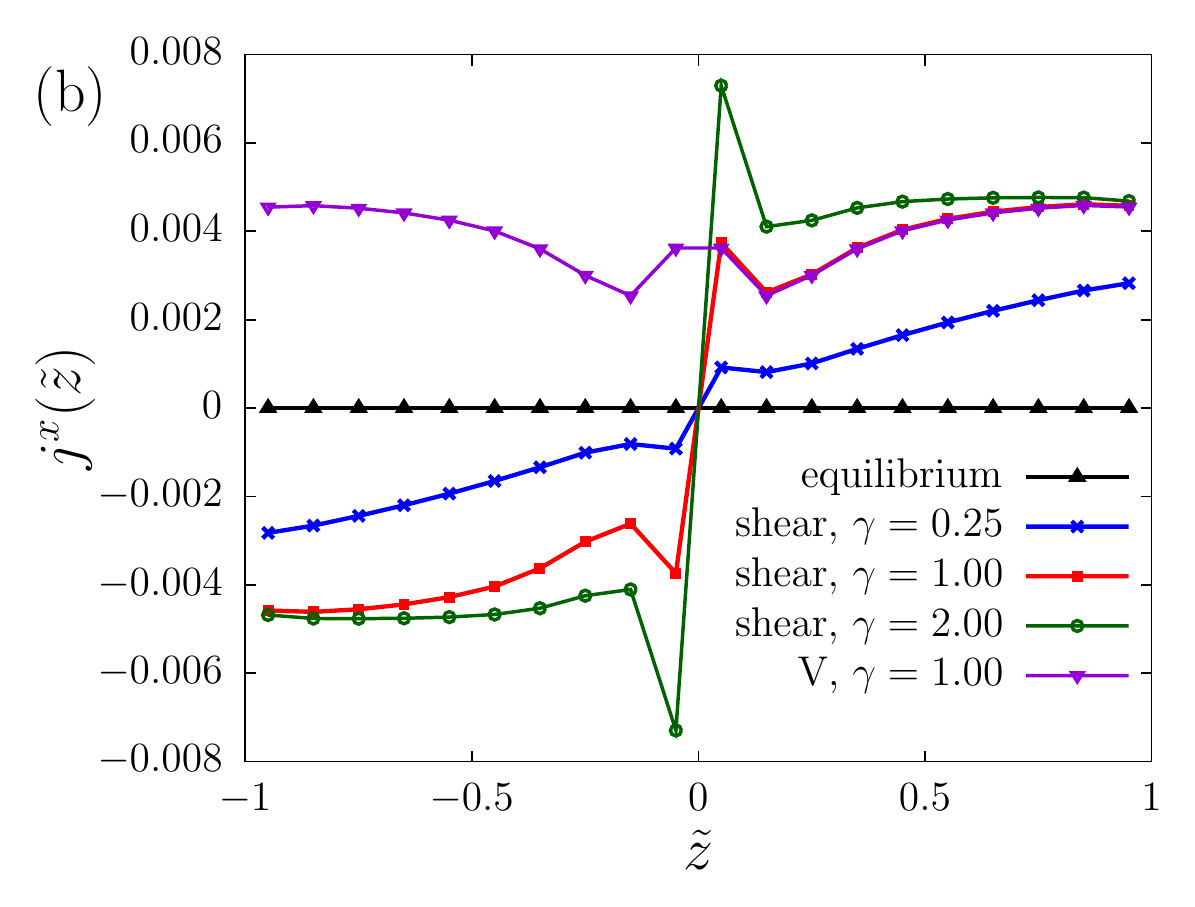}
\label{FigCurrent_Roughening}
}
\caption{(Color online) Order parameter current profile component $j^x(\tilde{z})$, for the system
parameters $L_x = 128$, $L_y = 128$, $L_z = 20$. (a) Temperature $T/T_c = 0.75$. Results are shown
for equilibrium (zero current), shear-like and V-shaped drive, and the case of mixed symmetry in the
driving field. In the latter case the lower and upper-half $\gamma$ values are $\gamma_l = 0.25$,
$\gamma_u = 0.5$. For $z>0$, the currents resulting from shear-like and V-shaped drive with
$\gamma=3$ coincide, as do those from mixed symmetry and shear with $\gamma = 0.5$. (b) $T=2.4$.
Results are shown for equilibrium, as well as shear-like and V-shaped drive with various values of
$\gamma$.
}
\label{FigCurrent}
\end{figure}

As shown in Fig.~\ref{FigCurrent}, the shear-like drive $F_{x}(z) = \gamma z$, $F_y(z) = 0$ gives a
purely odd order parameter current component $j^x(z)$. Since the $y$ component of the field is zero,
$j^y(z) = 0$. For the V-shaped drive, the current is an even function of $z$, since $F_x(z) = \gamma
\left|z\right|$. We thus expect transport along $x$ for the shear-like drive, but none for the
V-shaped drive. For the same value of $\gamma$, the currents for the two drive types almost coincide
in the region $z>0$, where the driving fields are the same in magnitude and direction.

In Figs.~\ref{FigCurrent_075} and \ref{FigCurrent_Roughening}, $j^x(z)$ is shown for various drive
gradients $\gamma$, for temperatures above and below the (equilibrium) roughening transition. Looking
at the high temperature data in Fig.~\ref{FigCurrent_075} we see that for small $\gamma$,
$\left|j^x(z)\right|$ has maxima at the walls; as $\gamma$ is increased, plateaus develop with the
maximum current shifted slightly from the wall. Eventually for strong drive the maxima become
localized near the interface.  This reflects the competing effects of local drive strength and
current carrier availability ($+-$ pairs): for large $\gamma$, the drive strength is essentially
saturated at the walls, so the greater carrier density at the interface eventually becomes more
important. Below the bulk roughening temperature ($T = 2.4$, Fig.~\ref{FigCurrent_Roughening}), the
current $\left|j^x(z)\right|$ also has maxima at the walls for small $\gamma$, and quickly develops
maxima at the middle two layers as $\gamma$ is increased. These maxima appear for much weaker drive
(approximately six times smaller $\gamma$) than they do for $T/T_c = 0.75$. They are also localised
to the two middle layers either side of the interface, and are much more pronounced than at the
higher temperature; this indicates that at low temperatures the interface region is very sharp,
reduced to approximately two lattice spacings. For strong drive, the greater carrier density at the
interface again `wins', and these maxima become global. We also note that $\left|j^x(z)\right|$ is
roughly five times smaller than that at the higher temperature, since the carrier density is much
smaller due to the increased bulk and interfacial order.

Finally, Fig.~\ref{FigCurrent_075} also shows an example of \emph{mixed} symmetry in the current
profile. The driving field is of the `V' type, but with different values of $\gamma$ in the upper
and lower halves of the system: $\gamma_l = 0.25$ in the lower half, $\gamma_u = 0.5$ in the upper.
Thus the total driving field can be written in the form $F(z) = \gamma_1 \left| z \right| + \gamma_2
z$, with $\gamma_1 = 0.75/2$, $\gamma_2 = 0.25/2$, showing the even and odd components explicitly.
The current profile reflects the asymmetry in the drive: in the lower half of the system, the
current matches that for a (symmetric) V-shaped drive with $\gamma = 0.25$, while in the upper half,
it matches that for either V or shear with $\gamma = 0.5$ (see Fig.~\ref{FigCurrent_075}) -- the
crossover occurs over a single lattice spacing.

\subsubsection{Space-time correlations.} \label{subsec:tcor}

\begin{figure}[ht]
\begin{center}
\includegraphics{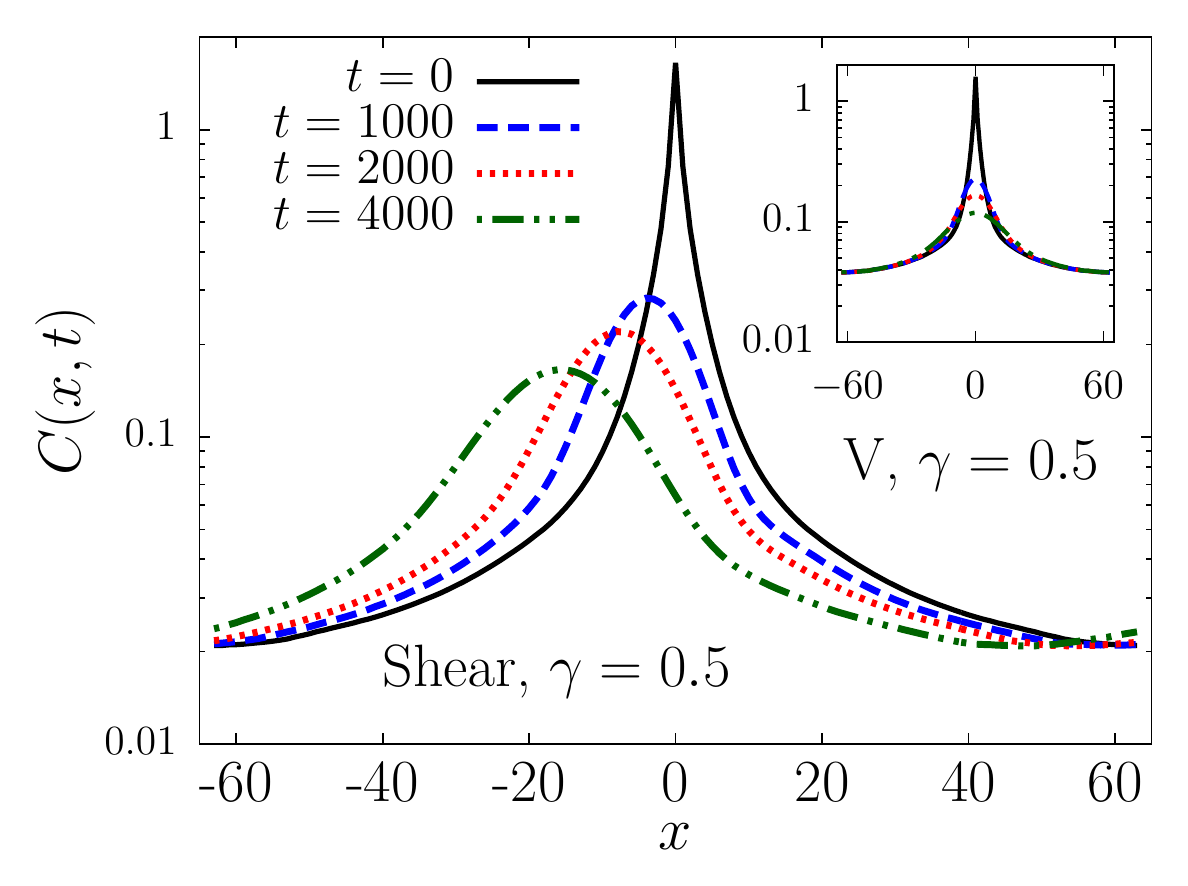}
\caption{(Color online) Time-displaced height-height correlation function $C(x,y=0,t)$ for a $128
\times 128 \times 10$ system at $T/T_c = 0.75$. Main panel: snapshots at time displacements $t$, as
indicated, for shear-like drive with gradient $\gamma = 0.5$. The peak of the correlation function
moves to the left with increasing time difference, indicating capillary wave motion. Inset: results
for V-shaped drive with strength $\gamma = 0.5$. Correlations decay without movement of the peak.}
\label{FigCxt}
\end{center}
\end{figure}

We investigate whether the conjecture for the occurrence of capillary wave motion holds for $3d$
systems by measuring $C(x,y=0,t)$ for different forms of driving field, which produce differing
current profiles. For current profiles with odd symmetry in $z$ (or more generally, profiles with an
odd \emph{component}), we expect to see evidence of capillary wave transport in $C(x,y=0,t)$.
Fig.~\ref{FigCxt} shows that this is indeed the case -- see the main panel for results for
$C(x,y=0,t)$ for shear-like drive, and the inset for V-shaped drive. For time difference $t=0$, the
peak lies symmetrically around $x=0$ due to the translational invariance ensured by the periodic
boundaries along $x$. However, at time differences $t>0$, the peak moves to negative $x$ values for
shear-like drive, indicating that now the greatest correlations are between spatially-displaced
points. We interpret this to mean that wave-like height fluctuations are being coherently
transported along the interface by the drive. For the V-shaped drive, the peak remains at $x=0$ for
all $t$, showing the absence of wave motion. In both cases, correlations decay with increasing time
difference, due to thermal noise. We note that the rate of decay of correlations is much faster for
driven systems than it is for equilibrium systems with Kawasaki dynamics.

We have also investigated other forms of driving field, for example spatially uniform driving field
in the $x$ direction, $F_x \equiv f = \textrm{const}$, $F_y = 0$ and step-like drive: $F_x(z) =
\textrm{sgn}(z)\cdot f$. The former produces an even order parameter current profile whereas the
latter  an odd one, and results for $C(x,0,t)$ (not displayed) show that wave motion does not occur
for a uniform drive but occurs for a step-like drive, consistent with the conjecture. For the case
of the asymmetric V-like drive discussed in the previous section, which has mixed symmetry, we
expect to see wave movement, since the current profile is not purely even, but like the driving
field itself, can be written as a sum of even and odd components. Indeed we find this is the case,
with the peak of $C(x,0,t)$ moving with time. From these results we conclude that the criterea for
capillary wave motion are the same in the $2d$ and $3d$ Ising systems.

\begin{figure}[ht]
\begin{center}
\includegraphics{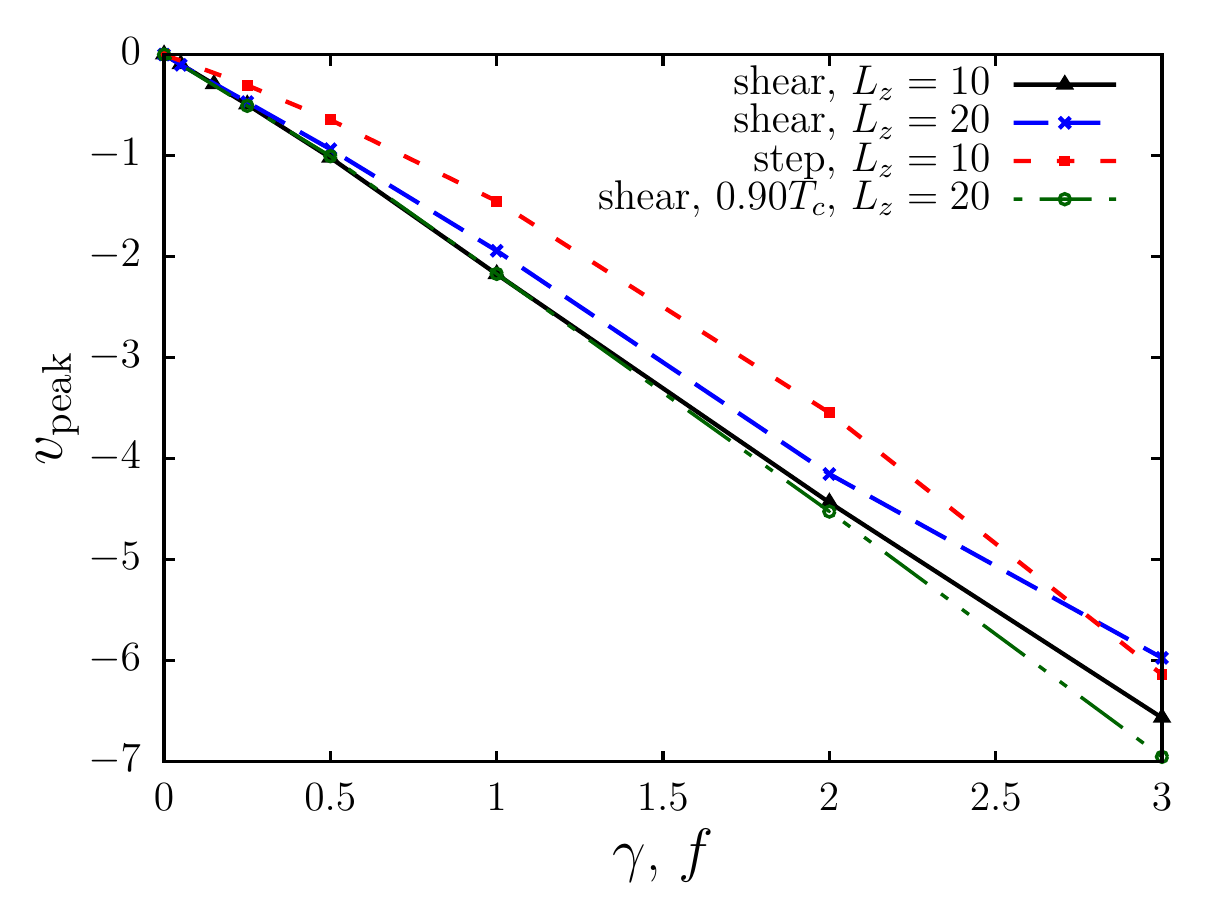}
\caption{(Color online) Speed of movement of the peak of $C(x,0,t)$, $v_{\rm peak}$, as a function
of driving strength, $\gamma$ or $f$, for shear- and step-like drives, respectively. Shown is the
effect of varying $L_z$ for fixed driving strength and temperature, and also the effect of
increasing the temperature from $T/T_c = 0.75$ to $0.9$.
}
\label{FigVPeak}
\end{center}
\end{figure}

Having established the occurence of wave motion, it is natural to investigate the dependence of the
wave velocity on system parameters. We measure the speed of the peak of $C(x,0,t)$, $v_{\rm peak}$,
and vary the driving strength ($\gamma$ for shear-like drive, $f$ for step-like drive), temperature
and wall separation $L_z$. As shown in Fig.~\ref{FigVPeak}, $v_{\rm peak}$ shows linear variation
with $\gamma$ or $f$ for fixed temperature and system size, for $\gamma, f \lesssim 2$. For
shear-like drive, the gradient of $v_{\rm peak}(\gamma)$ is close to 2 in this range. We also see
that varying $L_z$ has a rather small effect on the peak velocity -- doubling $L_z$ from 10 to 20
reduces the gradient of $v_{\rm peak}(\gamma)$ by only a few percent. Changing the temperature from
$0.75T_c$ to $0.90T_c$ also has small effect, in the other direction -- the peak moves faster for
the higher temperature at given $L_z$, $\gamma$. For the step drive, $v_{\rm peak}$ also seems to be
linear in driving strength $f$ for small $f$, with a reduced gradient compared to shear-like drive.
For both forms of drive, non-linearity appears to set in for $\gamma \gtrsim 2$. For the mixed
symmetry case, where the driving field could be written as $F(z) = \gamma_1 \left| z \right| +
\gamma_2 z$, we find that the velocity of motion is smaller than that for a purely odd field with
$\gamma = \gamma_2$ -- the velocity in the mixed case is linear in $\gamma_2$ and approximately 80\%
that of the pure case for $\gamma, \gamma_2 \lesssim 2$. For lower temperatures near and below the
equilibrium roughening temperature, we find that the interfacial motion still occurs, with a much
reduced velocity; correlations also decay much more quickly with time.

\subsubsection{Dispersion relation} \label{subsec:disprelat}

In order to characterize the dynamics of capillary waves we consider the evolution of the spatial 
Fourier modes ${\tilde h}({\bf k},t)$ of the height function $h({\bf r},t) := h(x,y,t)$ defined by:
\begin{equation}
\label{eq:fs}
h({\bf r},t)=\sum_{{\bf k}=1}^{{\bf L}}e^{2\pi i{\bf (k/L)}\cdot{\bf r}}{\tilde h}({\bf k},t),
\end{equation}
where  ${\bf k}=(k_x,k_y)$ and ${\bf r}=(x,y)$ are two-dimensional vectors and the division ${\bf
k/L}$ is defined as ${\bf k/L}=(k_x/L_x,k_y/L_y)$. (Recall that $x$ and $y$ are integer coordinates
of the lattice.) The sum denotes summations over integer  $k_x$ and $k_y$ from 1 to $L_x$ and $L_y$,
respectively. Because $h(x,y,t)$ is a real function, the Fourier transform has the conjugate
symmetry ${\tilde h}(k_x,k_y,t)={\tilde h}^{\ast}(L_x-k_x,L_y-k_y,t)$, i.e., there are $L_x/2 \times
L_y/2$ independent terms in \eqref{eq:fs}. Each complex Fourier component ${\tilde h}({\bf k},t)$
can be written in terms of its modulus and  phase $\phi({\bf k},t)$:
\begin{equation}
\label{eq:modphase}
{\tilde h}({\bf k},t) = \vert {\tilde h}({\bf k},t) \vert e^{\phi({\bf k},t)},
\end{equation}
where $-\pi \le \phi({\bf k},t) \le \pi$. If $\phi({\bf k},t) = 
\omega(k_x,k_y)t = (2\pi k_x/L_x)v_x(k_y)t$, each mode would correspond to a travelling wave 
moving in the $x$ direction with a velocity $v_x$ and $h(x,y,t)=h(x+v_xt,y)$ (with no dispersion).
\begin{figure}[ht]
\begin{center}
\includegraphics[scale=1]{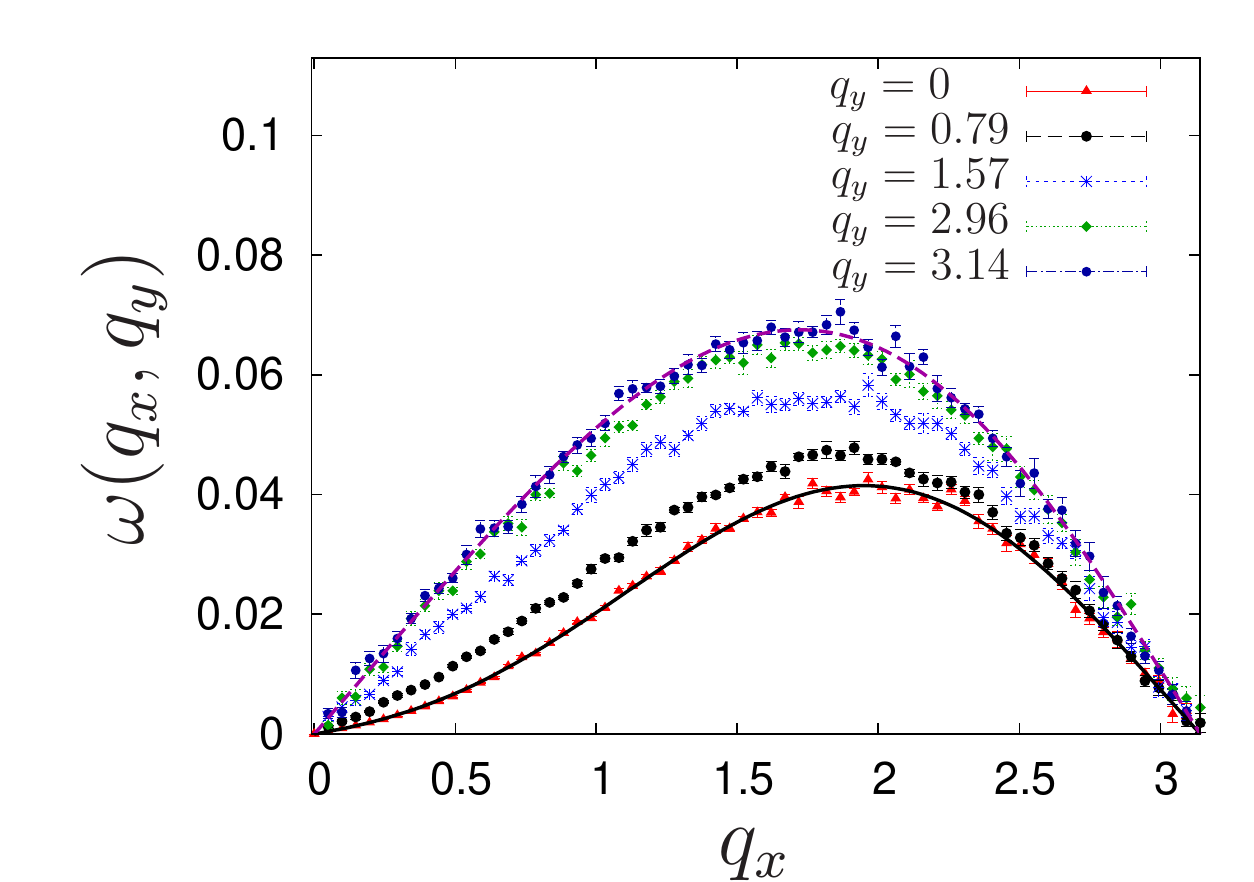}
\caption{(Color online) Dispersion relation $\omega(q_x,q_y)$  for shear-like drive as a function
of the wave number $q_x$ for several values of $q_y$. Data are for the system parameters $L_x =
128$, $L_y = 128$, $L_z = 10$ at  temperature  $T/T_c = 0.9$. Solid lines are fits to the analytical
formula \eqref{eq:fit}.}
\label{FigDisp}
\end{center}
\end{figure}
In a steady state, the phase shift of each mode is a fluctuating quantity, with a distribution which
when measured with increasing time intervals spreads and decays quickly to zero. However, at short
times, we are able to measure its mean value in unit time to obtain the dispersion relation of the
frequency $\omega$ as a function of the wave vector ${\bf q}=(q_x,q_y)=(2\pi k_x/L_x,2\pi k_y/L_y)$
as:
\begin{equation}
\label{eq:disp_def}
 \omega({\bf q})= \arg \langle \tilde
h^\ast({\bf q},t) \tilde h({\bf q},t+dt) \rangle/dt.
\end{equation}
Results for $\omega({\bf q})=\omega(q_x,q_y)$ calculated for $dt$ equal to $1/10$ of an MC sweep are
shown in Fig.~\ref{FigDisp} as a function of $q_x$ for several values of $q_y$. We observe similar
behaviour for all considered temperatures, including below the roughening temperature (data not
shown). The shape of  $\omega(q_x,0)$ is very similar to that obtained in the two-dimensional system
(see Ref.~\cite{MovPaper}), which suggests to use the  same analytical formula for describing the
dispersion relation:
\begin{equation}
\label{eq:fit}
\omega(q_x,q_y=const)=(v+2u)\sin(q_x)-u\sin(2q_x)+s\sin^2(q_x)
\end{equation}
with small-$q_x$ expansion 
\begin{equation}
\label{eq:fit1}
\omega(q_x,q_y=const)=vq_x+sq_x^2+(u-v/6)q_x^3. 
\end{equation}
Indeed, as can be seen in the Fig.~\ref{FigDisp}, our  data  for $\omega(q_x,0)$  fit well with
$v=0.0089(8), u=0.01095(15), s=0.0071(9)$.  Waves with larger  $q_y$ are less dispersive in the
sense that  the corresponding coefficients $u$ and $s$ are smaller. The fastest mode is the one with
$q_y=\pi$, or a wavelength of two lattice spacings; $\omega(q_x,\pi)$ can also be fitted using
(\ref{eq:fit}) with $v=0.0547(5), u=0.00530(26), s=0.0014(15)$. As we pointed out in
Ref.~\cite{MovPaper}, if the dynamics of the height function are modelled by the following linear
transport operator (propagator)
\begin{equation}
\label{eq:tranp_operator}
\hat L=(\partial_t -v\partial_x+u\partial^3_x),
\end{equation}
then the Ansatz in the form of the  travelling wave $A\exp(i(\omega t+{\bf q}\cdot {\bf r}))$ yields
the first and the second terms in the fit function \eqref{eq:fit}. In \eqref{eq:tranp_operator},
time is treated as a continuous variable whereas spatial derivatives are discrete: $\partial_x
h(x,y,t)=\left[h(x+1,y,t)-h(x-1,y,t)\right] / 2$, and a 5-point stencil is used for $\partial^3_x$.
The third term, which for small $q_x $ gives the quadratic dependence in \eqref{eq:fit1}, can be
obtained if one allows an imaginary contribution $i s(\partial_x^2+\partial_x^4/4)$ (with 3- and 5-
point stencils) to the linear operator $\hat L$. One may be able to understand the presence of
complex coefficients in the transport equation for the height function $h({\bf r},t)$, by
recognising that the plane wave solution above neglects the dependence of the amplitude $A$ on the
wave number. In fact the average modulus $\vert {\tilde h}({\bf q},t) \vert$ of each complex Fourier
component varies significantly with  $q=\vert {\bf q}\vert$, even in the absence of driving (see the
plot for the structure factor Fig.~\ref{FigStruct}). Taking this into account, it should be possible
to derive an equation for the amplitude as well as the phase -- this may aid in understanding the
presence of real and imaginary parts in the transport operator. The $t \to \infty$ limit of the
solution of the amplitude equation should yield the static structure factor, which may be compared
to simulation data. We leave the interesting and difficult problem of deriving the full transport
equation from these considerations to future work.

\section{Conclusions and outlook}
\label{sec:concout}

We have presented evidence from Monte Carlo simulations that the main characteristics of interfacial
structure and dynamics in the laterally driven stochastic lattice gas model are generic and persist
to three dimensions. Far above the equilibrium roughening temperature the structure of the interface
confined between two walls is affected by lateral driving in a way similar to that resulting from an
increase of confinement (by reducing the distance between the walls) of the \emph{equilibrium}
system. However, the effect of drive is much weaker in three dimensions than in two, plausible due
to the different nature  of low-energy interfacial fluctuations, i.e., the ``spike''-like
excitations rather then interface ``wandering''. In the case of the shear-like drive, our findings
for the decay of the height-height correlation functions and for the structure factor are in partial
agreement with recent results from fluctuating hydrodynamics \cite{ThieBick}. The discrepancy
concerns the behaviour in the vorticity direction. We have found a decrease of correlation length in
this direction whereas hydrodynamic calculations predict an increase. Also, our data show that the
structure factor is suppressed both in the drive and the vorticity directions. In
Ref.~\cite{ThieBick} it is concluded that $S(q_x,q_y)$ is unaffected in the vorticity direction.
Moreover, our results for the interfacial width are in agreement with the experiment of
Ref.~\cite{DerksShear}, but the results for the interfacial correlation length in the flow direction
are not. It would certainly be desirable to carry out more studies, experimental, theoretical and
simulation-based, to clarify these discrepancies. Nearer (and also below) the bulk roughening
temperature, the confinement/ordering effect of the drive is reduced, since there are no large
fluctuations to ``smoothen out''. The picture of an equilibrium system under a greater effective
confinement no longer seems to apply for these low temperatures.

The conjecture made in Ref.~\cite{MovPaper} for the occurence of lateral transport of interfacial
fluctuations is supported by our results in three dimensions. The transport also occurs for low
temperatures below the equlibrium roughening transition. However, the dynamics of the lateral
propagation of capillary waves which gives rise to the observed dispersion relation is still not
understood. A full treatment should include non-linear effects, and treat the time and wave-vector
dependence of the Fourier amplitude. Further insight, especially into the effect of coupling between
bulk and interfacial degrees of freedom, could be gained by considering theoretical models based on
the order parameter \cite{bray}.

%\bibitem{FiskWidom} S. Fisk and B. Widom, J. Chem. Phys. {\bf 50}, 3219 (1969).

\end{document}